\documentclass[journal]{IEEEtran}
\usepackage{cite}
\usepackage{amsmath,amssymb,amsfonts}
\usepackage{algorithmic}
\usepackage{graphicx}
\usepackage{textcomp}
\usepackage{subfigure}
\usepackage{booktabs} 
\usepackage{xcolor}

\newtheorem{remark}{Remark}
\newtheorem{problem}{Optimization Problem}

\newcommand{\figsize}{0.625}

\DeclareMathOperator*{\argmin}{argmin}

\begin{document}
\title{Balancing-Aware Charging Strategy For Series-Connected Lithium-Ion Cells: A Nonlinear Model Predictive Control Approach}
\author{Andrea Pozzi, Massimo Zambelli \IEEEmembership{Student, IEEE}, Antonella Ferrara \IEEEmembership{Senior Member, IEEE}, and Davide M. Raimondo
\thanks{Manuscript submitted January 11, 2019.}
\thanks{A. Pozzi, M. Zambelli, A. Ferrara and D.M. Raimondo are with the Dipartimento di Ingegneria Industriale e dell'Informazione, University of Pavia, 27100 Pavia, Italy (e-mail:{\tt\small \{andrea.pozzi03, massimo.zambelli01\}@universitadipavia.it, \{antonella.ferrara, davide.raimondo\}@unipv.it}).}}

\maketitle

\begin{abstract}
Charge unbalance is one of the key issues for series-connected Lithium-ion cells. Within this context, model-based optimization strategies have proven to be the most effective. In the present paper, an ad-hoc electrochemical model, tailored to control purposes, is firstly presented. Relying on this latter, a general nonlinear MPC   for  balancing-aware optimal charging is then proposed. In view of the possibility of a practical implementation, the concepts are subsequently specialized for an easily implementable power supply scheme. Finally, the nonlinear MPC approach is validated on commercial cells using a detailed battery simulator, with sound evidence of its effectiveness. 
\end{abstract}


\section{Introduction}
\label{sec:introduction}
\IEEEPARstart{I}{n} the last few decades, the ever increasing demand for technologies which rely on portable energy sources (e.g. electric and hybrid vehicles, consumer electronics, microgrids and IoT-related devices) has determined a growth in the request for high-performance energy accumulators. Among the possible manufacturing chemistries, Lithium-ion (Li-ion) batteries have proven to be the most promising \cite{Dunn2011,Tarascon2011,lu2013review}. In order to provide high terminal voltage, total capacity and available power, battery packs are usually composed of several cells, arranged in  series and parallel connections.
Due to unavoidable inconsistencies in the manufacturing process and non-homogeneities in the operating conditions (e.g. temperature and ageing effects, which appear over time), the cells composing  a battery pack exhibit slightly different features (e.g. in terms of internal impedance, self-discharge rate and physical volume)  \cite{Bentley1997,Dubarry2010}. As a consequence, appreciable unbalancing in the stored charge arises already after few charge/discharge cycles if conventional charging methods are employed \cite{Cao2008}. This latter is a limiting factor in terms of a complete battery exploitation (undercharging), state of health preservation (premature and non-uniform wear) and safety (overcharging and overdischarging phenomena). In particular, undercharging is a consequence of an early interruption of the charging procedure due to the reaching of the voltage threshold from a subset of cells. Premature and non-uniform degradation, instead, comes mainly from overvoltage exposures and high temperature spikes, which also affect the self-discharge rates. Moreover, during operation, temperature gradients arise between the different cells and even inside of each of them. All of these facts, in turn, contribute to an increase in the differences among the cells characteristics, which eventually lead to greater unbalance. 

In order to overcome such issues, a huge research effort has been made towards the development of algorithms implementing the idea of State of Charge (SOC) equalization. Two main categories can be identified for such balancing strategies, namely passive and active methods \cite{Moore2001, Wei2009, Daowd2011}. The former are based on energy dissipation, while the latter, in order to limit energy losses, rely mainly on charge redistribution . Among all of them, different structures can be adopted, e.g. cell bypass, cell-to-cell, cell-to-pack, pack-to-cell, cell-to-pack-to-cell \cite{Gallardo-Lozano2014}. 
 
At an earlier research stage, equalization was carried out using only model-independent low-level control strategies. Dissipative approaches have been proposed, e.g., in \cite{Bjork1986,Lindemark1991,Zheng2014}, while active voltage-based equalization methods have been presented, e.g., in \cite{Bonfiglio2009,Imtiaz2013,Phung2014,Kim2014}. 
Although the voltage can easily be measured in practice, some works have highlighted that a performance improvement can usually be achieved relying on capacity-based algorithms\cite{Einhorn2011, Caspar2014}, which however require an estimation of the state of charge, e.g. through observations \cite{Plett2004,Zhang2011,Waag2014,Bizeray2015}. This approach implies the need for a mathematical model of the cells, whose accuracy directly affects the overall algorithms performance. Several cell models have been proposed in the literature, which can mainly be  classified into two categories, namely Equivalent Circuit Models (ECMs) and Electrochemical Models (EMs). While those in the first set give  a simplified equivalent description, which only includes lumped states and parameters, the ones in the second group provide an accurate representation of the internal phenomena at the expense of heavier computational burden. 

The most used electrochemical model is the well-known Pseudo-Two-Dimensional (P2D) model \cite{Doyle1993,Doyle1995}, also referred as Doyle-Fuller-Newman (DFN) model, which consists of tightly coupled and highly nonlinear Partial Differential Algebraic Equations (PDAEs). Note that, due to their complexity, such PDAEs require proper advanced integration methods. Furthermore, it has been shown in the literature that some parameters of the DFN model are unidentifiable without invasive experiments \cite{Forman2012,Bizeray2016,Lopez2016} and that some observability issues \cite{Moura2015} are present. For these reasons, it appears evident that the P2D model is more suited for  simulation purposes rather than control and estimation. In order to preserve physical meaning and a sufficiently high accuracy while providing observability and identifiability, electrochemical reduced order models can be used \cite{Chaturvedi2010, DiDomenico2010, Moura2014, Pozzi2018}. For instance, the Single Particle Model (SPM) \cite{Ning2004,Santhanagopalan2006} and the Single Particle Model with Electrolyte dynamics (SPMe) \cite{Moura2017} are derived from the P2D considering each electrode as a single spherical particle. The resulting description is greatly simplified but still includes Partial Differential Equations (PDEs) related to the diffusion of the ions in the solid phase (and also in the electrolyte in case of SPMe), which partially preserve the original computational complexity.

% ALTRE CITAZIONI
Relying on a mathematical representation of the cells behaviour given by one of the available models, optimization-based balancing approaches can be developed.   Model Predictive Control (MPC, \cite{Maciejowski2002, Camacho2013}) is an optimization-based control technique which, due to   its ability to deal with complex systems in an optimal and effective way, considering also constraints, appears particularly suitable within the context of Lithium-ion batteries balancing.  For instance,  in  \cite{Danielson2012} balancing is carried out considering an arbitrary connection topology  for the cells, which are modeled as first-order integrators. A linear SOC model is employed by the authors in \cite{McCurlie2017} for minimum-time balancing of a battery pack. In \cite{Salamati2017} an adaptive MPC is proposed and a slightly more accurate model is considered in order to redistribute energy during the charging procedure. 
Even if considerable improvements have been shown in the literature, few issues are still evident. As a matter of fact, almost all the proposed works rely on relatively simple cell models. It appears then difficult to assess the validity of those algorithms that are tested only in simulation with such simple models considered also as the real plants. These latter are indeed expected to miss the representation of some key  features, at least in particular situations. For instance, temperature as well as ageing dynamics are often neglected, even if they lead to performance degradation over time. In some high-power applications, e.g. electric and hybrid vehicles, cooling systems are arranged in order to dissipate the excessive heat. In such cases, thermal couplings between cells and between cells and coolant play an important role which should be accounted for in the equalization procedure. 

The majority of the equalization algorithms are designed to operate offline, at time instants defined by the specific policies, with the aim to mitigate the unbalancing effects of different cells characteristics. This seems hardly applicable in a practical scenario since it usually requires an idle time in which only the charge balancing is performed (except for the cases in which it is performed during the discharge and the load current profile is known in advance, as shown e.g. in \cite{Altaf2017}). Furthermore, the concept of SOC balancing is effective in maximizing the battery utilization if one supposes that the total capacity and the Solid-Electrolyte Interphase (SEI) resistance are always equal among the cells throughout the battery lifetime. In a real framework, however, due to all the previously mentioned phenomena, this is never the case. Therefore, an offline algorithm such as that usually utilized, not only is unable to maximize battery exploitation, but can also lead to overcharge in the successive charging cycle if a standard method is adopted (e.g. Constant Current - Constant Voltage, CC-CV). In fact, if one balances the SOC at a value lower than the maximum, right before a charge, then the SOCs will diverge one from each other due to the difference in the capacities and the application of a unique current. The only reasonable solution in such cases appears to be the so-called top balancing \cite{Barsukov2009}, which consists in balancing the cells SOC to their maximum value at the end of the charge.

In the present paper a general charging strategy based on Nonlinear MPC (NMPC) is proposed with the aim of overcoming the mentioned problems. Then, it is specialized for the case in which each cell is embedded into a power unit (PU) along with two switches, which allow for the cell bypass. Starting from the SPMe, an even more simplified description suitable for control purposes is derived, based only on Ordinary Differential Equations (ODEs). Beside a lumped internal temperature model, already present in the literature \cite{Guo2011,Perez2016,Campbell2019}, the thermal couplings between the cells composing the pack is modelled for an arbitrary topological disposition. Furthermore, in order to enhance the description, coolant temperature as well as ageing dynamics are also considered. This latter accounts for capacity losses and film resistance growth over time, whereas the former can be particularly important in automotive applications. Considering all the features mentioned above in a unique formulation represents a novelty to the best of authors knowledge.

In addition, nullifying the unbalance at the end of each charge cycle, the proposed approach is designed to be able to replace the presence of an equalization system coupled with the battery pack. Due to the simple structure of the power supply control scheme adopted for the practical implementation proposal, which only consists in a variable current generator and a set of PUs, energy losses are maintained at a reasonable level. Furthermore, the predictive nature of the algorithm makes it possible to avoid redistribution, thus increasing energy efficiency and reducing charging time. Suitably designed constraints are enforced for both safety and battery health preservation reasons. The effectiveness of the presented strategy is validated in simulation using an extended version (available on request) of the LIONSIMBA toolbox \cite{Torchio2016}, which is a numerical implementation of the P2D. The simulation environment is set considering a full order model for the solid diffusion law and a lumped model for the thermal dynamics. In order to consider a real world scenario the cells parameters are taken from \cite{Ecker2015,Ecker2015a}, where a complete parametrization of the Kokam SLPB 75106100 cell is presented.

With the exception of the temperature, the internal states of each cell are not measurable in practice. Nevertheless, in this paper the assumption of availability of all the relevant states is made. The development of a suitable observer and its coupling with the presented control system goes beyond the scope of this paper and will be investigated in the future.
This work focuses only on series-connected cells, similarly to what done in the literature. Research effort will be devoted to extend the applicability also to parallel-connected series of cells. 

\subsection{Main Contributions}
\label{sub:maincontrib}
The main contributions are here summarized for the readers convenience. First of all, a simplified electrochemical model which is suitable for online control but also able to describe all the relevant aspects of a Lithium-ion battery pack is formulated. In particular, beside the internal dynamics of each cell, the introduced model includes coupled thermal dynamics (consisting also of the cooling system) as well as ageing effects. Basing on this description, a general NMPC formulation is proposed for an optimal charging strategy able to achieve balancing at the end of each charge (top balancing). Additionally, the procedure is applied to a proposed supply scheme, easy to implement but more suited for mixed-integer optimization methods. Due to the superior scalability of continuous optimization over the discrete one, particular adaptations are proposed for the coupling of the scheme with the NMPC controller.

\subsection{Outline}
\label{sub:outline}
The remaining of the paper is organized as follows. In Section \ref{sec:model} a detailed explanation of the model components is given. In  \ref{sub:singlecell} the simplified version of the SPMe  for the single cell electrochemical behaviour is introduced, along with ageing and lumped thermal models. In  \ref{sub:thermal} the thermal couplings are described through a structure able to account for arbitrary relationships between the elements. Then, the model performance is assessed in  \ref{sub:validation}, through a sinusoidal charge-discharge input current. In Section \ref{sec:balancing}, the series balancing-aware optimal charge problem is formally stated (\ref{sub:optim}) and applied to the proposed power supply circuit (\ref{sub:implem}). Section \ref{sec:simu} is devoted to the evaluation of the proposed work. In particular, the testbed is described in  \ref{sub:testbed}, where the  characteristics of the considered cells and the battery pack structure are presented. All the cells exhibit different initial conditions in terms of capacity, SEI resistance and SOC, in order to emulate a real scenario in which the non-homogeneities  have led to an unbalanced situation. Firstly, in \ref{sub:cccv}, a standard charging protocol (namely, the CC-CV) is applied in order to  show the effects of such an initial mismatch. Then, a simple method that is based on the proposed circuit but only exploits voltage measurements is tested in \ref{sub:ray} so as to 
highlight the need for a model-based approach in order to achieve high performance. Eventually, the effectiveness of the proposed strategy is  made evident in \ref{sub:results}. Conclusions are finally drawn in Section \ref{sec:conclusions}, where possible future enhancements are also discussed.

\section{Battery Model}
\label{sec:model}
In this section, a control-oriented electrochemical model of the battery pack with thermal and ageing dynamics is presented, which is sufficiently accurate but also suitable for online control applications. In  \ref{sub:singlecell} the electrochemical behaviour of a single cell is considered along with the respective temperature dynamics and  ageing effects. The thermal coupling between cells and coolant is modelled in  \ref{sub:thermal}, where an arbitrary topology of the elements inside the battery pack is considered. At the end (\ref{sub:validation}), a sinusoidal charge-discharge cycle is performed to give an idea of the general accuracy of the model with respect to the P2D implemented in LIONSIMBA.

\subsection{Model of a Single Cell}
\label{sub:singlecell}
The SPMe, proposed in \cite{Moura2017}, provides an approximation of the DFN full order model which still relies on PDEs for describing  the diffusion of solid (through the Fick's law) and electrolyte concentrations. In this paper, we consider a simplified version of the SPMe in which the Fick's law is reduced for each electrode to an ODE by considering a polynomial approximation of the solid concentration along the particle radius \cite{Subramanian2005}. Furthermore, the electrolyte diffusion PDE is spatially discretized according to the Finite Volume (FV) method \cite{Eymard2000}, as already done in a similar context in, e.g., \cite{Torchio2016}. In addition, starting from the works in \cite{Ramadass2004, Torchio2016a} we augment the adopted model with the cell ageing effects in terms of SEI resistance growth and capacity fade, which are among the main degradation phenomena for a Lithium-ion cell \cite{Ramadass2003}. Note that, in the proposed model, some of the parameters exhibit a temperature dependency. In this work, this latter will be modelled according to the Arrenhius law
\begin{align}
\label{eq:arrhenius}
\psi(T(t))=\psi_0 e^{\frac{-E_{a,\psi}}{RT(t)}}
\end{align}
where $T(t)$ is the temperature, $E_{a,\psi}$ is the activation energy associated with the generic parameter $\psi(T(t))$ and $R$ is the universal gas constant. Such relationship will be explicitly highlighted for each of the temperature-dependent parameter. 

In this Subsection, the index $j \in \{p,s,n\}$ will refer to the three cell sections (respectively, cathode, separator and anode), while $i \in \{p,n\}$ will be used in equations valid only for the electrodes.

\subsubsection{Solid Concentration Dynamics}
\label{sub:solid}
According to the authors in \cite{Subramanian2005}, the lithium concentration profile along the radial coordinate  can be approximated as a fourth-order polynomial function of $r$, $0\leq r\leq R_{p,i}$, with $R_{p,i}$ being the particle radius. The parameters involved in this approximation depend only on the  average 
stoichiometry $\bar \theta_i(t)$ and the volume-averaged concentration flux $\bar q_i(t)$. The dynamics of these latter can be described in terms of the ODEs recalled below.

The average stoichiometry $\bar \theta_i(t)$ is defined, for each electrode, as the ratio between the average solid concentration $\bar c_{s,i}(t)$ and the  maximum  solid concentration $c_{s,i}^{max}$, i.e. 
\begin{align}
\bar \theta_i(t)=\frac{\bar c_{s,i}(t)}{c_{s,i}^{max}}
\end{align}

In order to minimize the number of ODEs required to describe the lithium solid concentration within each cell, we express the anodic stoichiometry in terms of the cathodic one as in \cite{DiDomenico2010}
\begin{align}
\bar \theta_n(t)= \theta^{0\%}_n+\frac{\bar \theta_p(t)-\theta^{0\%}_p}{\theta_p^{100\%}-\theta^{0\%}_p} (\theta_n^{100\%}-\theta^{0\%}_n )
\end{align}
where  $\theta^{100\%}_i$ and $\theta^{0\%}_i$ are the values of the stoichiometry respectively at the battery charged and discharged states (provided in the cells data-sheet). This relationship is guaranteed by the fact that the moles of lithium in the solid phase are preserved \cite{Klein2013}. 
%Taking into account also the coulombic efficiency $\mu_c(t)$,
The dynamics of the average stoichiometry  for the cathode can be expressed as
\begin{align}\label{eq:theta_dynamics}
\dot{\bar \theta}_p(t)=-\frac{\Delta \theta_p}{C(t)}I_{app}(t)
\end{align}
where $C(t)$ is the available capacity of the cell, $\Delta \theta_i= \theta_i^{100\%}-\theta_i^{0\%}$, and $I_{app}(t)$ is the applied current. Note that, the convention adopted in this work is such that the charging current is negative, i.e. $I_{app}(t) <0$.
The volume-averaged concentration flux can be described by 
\begin{align}
	\dot {\bar q}_i(t) = -30 \frac{D_{s,i}(T(t))}{R_{p,i}^2}\bar q_i(t) - \frac{45\Delta \theta_i c_{s,i}^{max}}{6R_{p,i}C(t)}I_{app}(t)
\end{align} 
where  $D_{s,i}(T(t))$ is the temperature-dependent solid diffusion coefficient. Then,  the positive and negative surface stoichiometries are given respectively by 
\begin{align}
\label{eq:stoichiometric}
\theta_i(t)= \bar \theta_i(t)+\frac{8R_{p,i}\bar q_i(t)}{35c_{s,i}^{max}}-\frac{R_{p,i}^2 \Delta \theta_i}{105D_{s,i}(T(t))C(t)}I_{app}(t) 
\end{align}
Finally, the normalized SOC $z(t)$ is defined as follows 
\begin{align}\label{eq:soc}
z(t)=\frac{\bar \theta_{n}(t)-\theta_n^{0\%}}{\theta_n^{100\%}-\theta_n^{0\%}}
\end{align}
Note that Equation \eqref{eq:soc} relies on the widely used convention that the cell SOC corresponds to the anodic one. Moreover, the presented equations are formulated in order to keep \eqref{eq:theta_dynamics} intuitive and simplifying a bit the notation. In particular, they are derived from the standard SPM description by taking the active material volume fraction $\epsilon^{act}_i(t)$   as a function of  the available capacity of the cell $C(t)$, i.e.
\begin{align}
\epsilon^{act}_p(t)&=-\frac{C(t)}{\Delta \theta_p A F L_p c_{s,p}^{max}}\\
\epsilon^{act}_n(t)&=\frac{C(t)}{\Delta \theta_n A F L_n c_{s,n}^{max}}
\end{align}
where $A$ is the contact area between solid and electrolyte phase, $F$ is the Faraday constant and $L_i$ is the layer thickness.  This definition is  chosen according to the fact that the SOC-OCV (Open Circuit Voltage) curve variations are small over the battery lifetime \cite{Waag2014}.

\begin{remark}
The polynomial approximation used above to describe the lithium solid concentration along the particle radial coordinate  can be replaced by a more accurate one, if needed. For instance, a spatial discretization using Chebyshev orthogonal collocation can be employed \cite{Bizeray2015}. 
\end{remark}

\subsubsection{Electrolyte Concentration Dynamics}
\label{sub:electrolyte}
The PDEs governing the  diffusion of the electrolyte concentration $c_{e,j}(x,t)$ in the SPMe \cite{Moura2017}  are the following
\begin{subequations}
\label{eq:ce_pde}
\begin{align}
\begin{split}
\epsilon_p \frac{\partial  c_{e,p}(x,t)}{\partial t}=&  \frac{\partial}{\partial x}\left[D_{e,p}^{eff}(T(t)) \frac{\partial c_{e,p}(x,t)}{\partial x} \right] \\ &-\frac{1-t_+}{FAL_p}I_{app}(t)\end{split}
\\
\epsilon_s \frac{\partial  c_{e,s}(x,t)}{\partial t}=&  \frac{\partial}{\partial x}\left[D_{e,s}^{eff}(T(t)) \frac{\partial c_{e,s}(x,t)}{\partial x} \right] \\
\begin{split}\epsilon_n \frac{\partial  c_{e,n}(x,t)}{\partial t}=&  \frac{\partial}{\partial x}
\left[D_{e,n}^{eff}(T(t)) \frac{\partial c_{e,n}(x,t)}{\partial x} \right] \\&+\frac{1-t_+}{FAL_n}I_{app}(t)
\end{split}
\end{align}
\end{subequations}
where $D_{e,j}^{eff}(T(t)) = D_e(T(t))\epsilon_j^{p_j}$ with $\epsilon_j$ the material porosity, $p_j$ the Bruggeman coefficient, $t_+$ the transference number and $D_e(T(t))$ the temperature-dependent diffusion coefficient within the electrolyte. The  boundary conditions are specified by the following equations \cite{Moura2017}

\begin{subequations}
\begin{align}\label{eq:boundary}
\frac{\partial c_{e,p}(0_p,t)}{\partial x}=&\frac{\partial c_{e,n}(L_n,t)}{\partial x}=0\\
D_{e,p}^{eff}(T(t))\frac{\partial c_{e,p}(L_p,t)}{\partial x}=&D_{e,s}^{eff}(T(t))\frac{\partial c_{e,s}(0_s,t)}{\partial x}\\
D_{e,s}^{eff}(T(t))\frac{\partial c_{e,s}(L_s,t)}{\partial x}=&D_{e,n}^{eff}(T(t))\frac{\partial c_{e,n}(0_n,t)}{\partial x}\\
c_{e,p}(L_p,t)=&c_{e,s}(0_s,t)\\
c_{e,p}(L_n,t)=&c_{e,n}(0_n,t)
\end{align}
\end{subequations}
where $0_j$ and $L_j$ are, respectively, the beginning and the end of the $j$-th section along the $x$-axis.

In this work, the spatial domain is divided for each section into $M$ non-overlapping volumes with centered nodes. The $k$-th volume, with $k=1, \cdots, M$, of the $j$-th section is centred at the spatial coordinate $x_{j,k}$ and spans the interval $\Omega_{j,k}=\left[x_{j,\bar{k}},x_{j,\underline{k}}\right]$, whose  width is  $\Delta x_j=L_j/M$. The PDEs are then discretized according to the FV method. Defining $c_{e,j}^{[k]}(t)$ as the average electrolyte concentration over the $k$-th volume of $j$-th section,  one has 
\begin{subequations}
\label{eq:ce}
\begin{align}
\begin{split}
\epsilon_p \frac{\partial  c^{[k]}_{e,p}(t)}{\partial t}=&  \left.\left[\frac{\tilde{D}_e(x,T(t))}{\Delta x_p}   \frac{\partial c_{e,p}(x,t)}{\partial x} \right]\right|_{x_{j,\underline{k}}}^{x_{j,\bar{k}}} \\&-\frac{1-t_+}{FAL_p}I_{app}(t)\end{split}\\
\epsilon_s \frac{\partial  c^{[k]}_{e,s}(t)}{\partial t}=&  \left.\left[\frac{\tilde{D}_e(x,T(t))}{\Delta x_s}   \frac{\partial c_{e,s}(x,t)}{\partial x} \right]\right|_{x_{j,\underline{k}}}^{x_{j,\bar{k}}}\\
\begin{split}
\epsilon_n \frac{\partial  c^{[k]}_{e,n}(t)}{\partial t}=& \left.\left[\frac{\tilde{D}_e(x,T(t))}{\Delta x_n}   \frac{\partial c_{e,n}(x,t)}{\partial x} \right]\right|_{x_{j,\underline{k}}}^{x_{j,\bar{k}}} \\&+\frac{1-t_+}{FAL_n}I_{app}(t)
\end{split}
\end{align}
\end{subequations}
where the terms are evaluated as explained in details in \cite{Torchio2016}. In particular, the electrolyte diffusion coefficients are computed as follows
\begin{align}
\tilde{D}_e(x,T(t))=
\begin{cases} D_{e,1} &\mathrm{if} \quad x \in \{x_{p,\underline{M}},x_{s,\overline{1}}\}\\
D_{e,2} \quad  &\mathrm{if} \quad x \in \{x_{s,\underline{M}},x_{n,\overline{1}}\}\\
D^{eff}_{e,j}(T(t)) &\mathrm{otherwise}
\end{cases}
\end{align}
with
\begin{subequations}
\begin{align}
D_{e,1}&=\mathcal{H}\left(D^{eff}_{e,p}(T(t)),D^{eff}_{e,p}(T(t)),\Delta x_p,\Delta x_s\right)\\ 
D_{e,2}&=\mathcal{H}\left(D^{eff}_{e,s}(T(t)),D^{eff}_{e,n}(T(t)),\Delta x_s,\Delta x_n\right) 
\end{align}
\end{subequations}
where $\mathcal{H}$ is the harmonic mean operator, defined as
\begin{align}
\mathcal{H}\left(\rho_1,\rho_2,\lambda_1,\lambda_2\right)=\frac{\rho_1 \rho_2 (\lambda_1 + \lambda_2)}{\rho_1\lambda_2+\rho_2\lambda_1}
\end{align} 
\subsubsection{Ageing Effects}
\label{sub:ageing}
The main consequences of ageing in Lithium-ion cells are a loss of capacity and a growth of the SEI film resistance \cite{Ramadass2003}. Such effects are primarily caused by high cycling currents and temperatures. In order to describe these degradation phenomena, we modify the models in \cite{Ramadass2004, Torchio2016a} to match the SPMe dynamics. In particular, the evolution of the residual available capacity $C(t)$ and the SEI resistance $R_{sei}(t)$ is as follows
\begin{subequations}
\begin{align}
	\dot C(t) &= \frac{3C(t)}{R_{p,n}A \Delta \theta_n c_{s,n}^{max}}\bar j_{side}(t) \label{eq:dotQ} \\
	\dot R_{sei}(t) &= -\frac{M_w}{\rho_n\nu}\bar j_{side}(t) \label{eq:dotrsei}
\end{align}
\end{subequations}
where $M_w$ is the molar weight of the negative electrode, $\rho_n$  the material density and $\nu$  the admittance of the film. The average side reaction flux is
\begin{align}\label{eq:jside}
	\bar j_{side}(t) = -\frac{i_{0,side}(t)}{F}e^{\left ( {\frac{0.5F}{RT(t)}\bar \eta_{side}(t)} \right )}
\end{align}
while the side reaction exchange current $i_{0,side}(t)$ is
\begin{align}
	i_{0,side}(t) = 
	\begin{cases} i_{0,base}\left ( \frac{\vert I_{app}(t) \vert}{I_{1C}}	\right )^w &\mathrm{if} \quad I_{app}(t)\leq 0\\
0 &\mathrm{otherwise}
\end{cases}
\end{align}
where $i_{0,base}$ is the base-side reaction current, $I_{1C}$ is the current corresponding to a 1C rate for the considered cell and $w$ is an empirically obtained coefficient. Note that no degradation occurs when the cell is discharged (i.e. for $I_{app}(t)>0$).
The side reaction overpotential $\bar \eta_{side}(t)$ is given by
\begin{align}
\bar \eta_{side}(t) = \bar \eta_{n}(t) + \bar U_n(t) - U_{SEI}
\end{align}
where $U_{SEI}$ is the reference potential for the SEI side reaction, while $\bar U_{n}(t)$ and $\bar \eta_n(t)$  are the negative Open Circuit Potential (OCP) and overpotential, respectively. In particular, the OCPs, for both electrodes, can be represented as  nonlinear functions of the surface stoichiometric coefficients  with expressions dependent on the particular considered cell (see \ref{sub:testbed}). On the other side, $\bar \eta_i(t)$ is given by
\begin{align}\label{eq:eta}
	\bar \eta_i(t) = \frac{2RT(t)}{F}\sinh^{-1} \left(\frac{\Delta \theta_i F R_{p,i}}{6\bar i_{0,i}(t)C(t)}I_{app}(t)			\right)  
\end{align}
with
\begin{align}
	\bar i_{0,i}(t) = Fk_i(T(t))\sqrt{\bar c_{e,i}(t)\theta_i(t)(1-\theta_i(t))}
\end{align}
where $k_i(T(t))$ is the temperature-dependent rate reaction constant, and $\bar c_{e,i}(t)$ is the average electrolyte concentration in the $i$-th section, approximated as follows
\begin{align}
\bar c_{e,i}(t)=\frac{1}{M}\sum_{k=1}^M c_{e,i}^{[k]}(t)
\end{align} 
 
\subsubsection{Output Voltage}
\label{sub:voltage}
\begin{figure}[!tb]
\begin{center}
\includegraphics[width=\figsize\columnwidth]{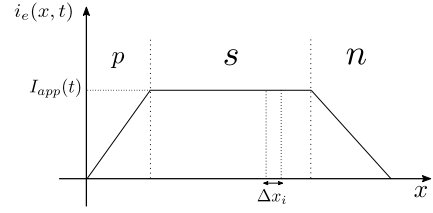}
\caption{Trapezoidal approximation of the ionic current.}\label{fig:ionic_current_trapezoidal}
\end{center}
\end{figure}
The terminal voltage is given by
\begin{align}
\label{eq:voltage}
\begin{split}
	V(t) =& \bar U_p(t) - \bar U_n(t) + \bar \eta_p(t) - \bar \eta_n(t) + \\
	&+ \Delta\Phi_e(t) + I_{app}(t)R_{sei}(t)
\end{split}
\end{align}
with elements defined according to equations \eqref{eq:dotrsei}, \eqref{eq:eta}, and
\begin{align}
	\Delta\Phi_e(t) = \Phi_e^{drop}(t) + \frac{2RT(t)}{F}(1-t_+)\log{\left(\frac{c_{e,p}^{[1]}}{c_{e,n}^{[M]}}	\right)}
\end{align}
in which $\Phi_e^{drop}(t)$ is approximated according to the assumption that the ionic current $i_e(x,t)$ has a trapezoidal shape over the spatial domain \cite{Moura2017} (see Fig. \ref{fig:ionic_current_trapezoidal} for a schematic representation), i.e.
\begin{align}
\Phi_e^{drop}(t)\simeq  -\frac{I_{app}(t)}{2 M} \left( \phi_p(t) + 2\phi_s(t) + \phi_n(t) \right)
\end{align}
In view of the mentioned assumption one can easily derive 
\begin{subequations}
\label{eq:phi}
\begin{align}
\phi_p(t)&=\Delta x_p \sum_{k=1}^M \frac{2k-1}{\kappa(c_{e,p}^{[k]}(t),T(t))\epsilon_p^{p_p}} \\
\phi_s(t)&=\Delta x_s \sum_{k=1}^{M} \frac{1}{\kappa(c_{e,s}^{[k]}(t),T(t))\epsilon_s^{p_s}}\\
\phi_n(t)&=\Delta x_n \sum_{k=1}^{M} \frac{2M-2k+1}{\kappa(c_{e,n}^{[k]}(t),T(t))\epsilon_n^{p_n}}
\end{align}
\end{subequations}
where $\kappa(c_{e,j}^{[k]}(t),T(t))$ is the temperature-dependent electrolyte conductivity for the $k$-th volume of the $j$-th section, which is usually expressed as an empirically derived nonlinear function of the electrolyte concentration in that volume (see \ref{sub:testbed}). Such relationship can in principle differ from cell to cell. Note that, the expressions in \eqref{eq:phi} are computed considering different electrolyte conductivity values for each one of the discrete volumes constituting the cell sections. This approximantion provides better accuracy with respect to e.g. \cite{Moura2017}, in which the electrolyte conductivity is assumed constant over the spatial domain.

\subsection{Battery Thermal Model}
\label{sub:thermal}
With a slight abuse of notation, from now on the variables $i$ and $j$ are used to indicate specific cells inside the battery pack. This latter is composed of $N$ cells, which exchange heat through direct contact with other cells (conduction) and possibly with a coolant fluid (by convection). Both conduction and convection phenomena are approximated using lumped thermal resistance parameters $R_{i,j} = R_{j,i}$, which are associated with the couple of cells $(i,j)$, and $R_{i}^{sink}$ that considers the interactions of the $i$-th cell with the coolant. As a matter of fact, any arbitrary disposition of the cells can be fully described from a thermal point of view through the choice of the resistance parameters. Having defined $C_{th,i}$ and $C_{th,sink}$ as the thermal capacities of the $i$-th cell and the sink respectively, the temperature dynamics of  the $i$-th cell is given by 
\begin{align}\label{eq:temperature}
C_{th,i} \dot{T}_{i}(t) = Q_i(t)-\frac{\Delta T_{i}^{sink}(t)}{R_{i}^{sink}} + \sum_{j \neq i} \frac{\Delta T_{j,i}(t)}{R_{i,j}}
\end{align}
where $T_i(t)$ and $T_{sink}(t)$ are the temperatures of the $i$-th cell and the coolant, $\Delta T_{j,i}(t) = T_j(t) - T_i(t)$ and $\Delta T_{i}^{sink}(t) = T_i(t) - T_{sink}(t)$.
The nonlinear term $Q_i(t)$ is the heat generation due to polarisation inside the cell, defined as
\begin{align}
\label{eq:qi}
Q_i(t)=\vert I_{app,i}(t) \vert \cdot \vert V_{i}(t) - (\bar U_{p,i}(t)-\bar U_{n,i}(t)) \vert 
\end{align}
where the terms in \eqref{eq:qi} are the instantaneous current, voltage and open circuit potentials associated with the $i$-th cell. The temperature dynamics of the sink is instead given by
\begin{align}
C_{th,sink} \dot{T}_{sink}(t) = \sum_{i} \frac{\Delta T_{i}^{sink}(t)}{R_{i}^{sink}} - \xi(t)
\end{align}
where $\xi(t)$ is the thermal power released by the cooling system to the external environment.

\subsection{Performance Assessment}
\label{sub:validation}
\begin{figure}[!tb]
\begin{center}
\includegraphics[width=\figsize\columnwidth]{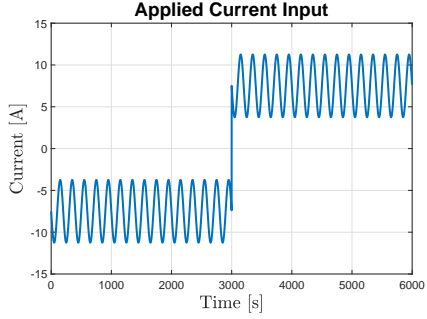}
\caption{Input sinusoidal charge-discharge current.}
\label{fig:input_validation}
\end{center}
\end{figure}
\begin{figure}[!tb]
\begin{center}
\includegraphics[width=\figsize\columnwidth]{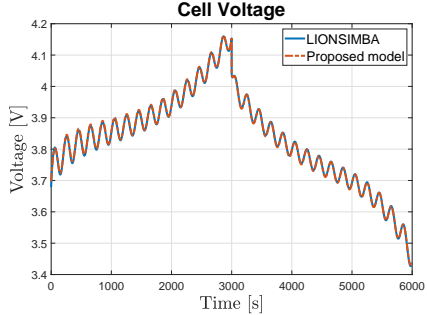}
\caption{Voltage comparison.}
\label{fig:voltage_validation}
\end{center}
\end{figure}
\begin{figure}[!tb]
\begin{center}
\includegraphics[width=\figsize\columnwidth]{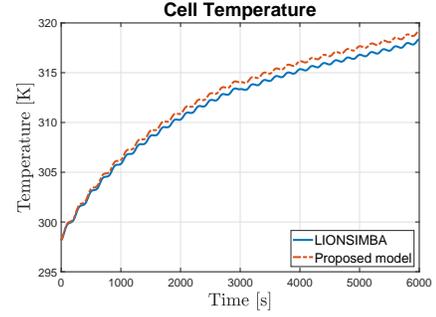}
\caption{Cell temperature comparison.}
\label{fig:temp_validation}
\end{center}
\end{figure}
\begin{figure}[!tb]
\begin{center}
\includegraphics[width=\figsize\columnwidth]{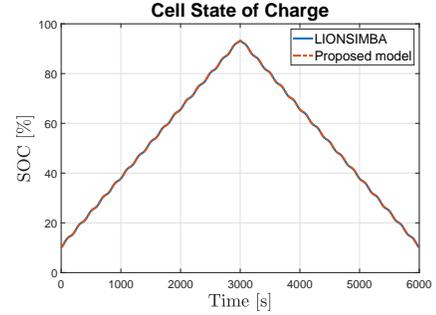}
\caption{SOC comparison.}
\label{fig:soc_validation}
\end{center}
\end{figure}
\begin{figure}[!tb]
\begin{center}
\subfigure[Cell Loss Capacity]{
\includegraphics[width=0.47\columnwidth]{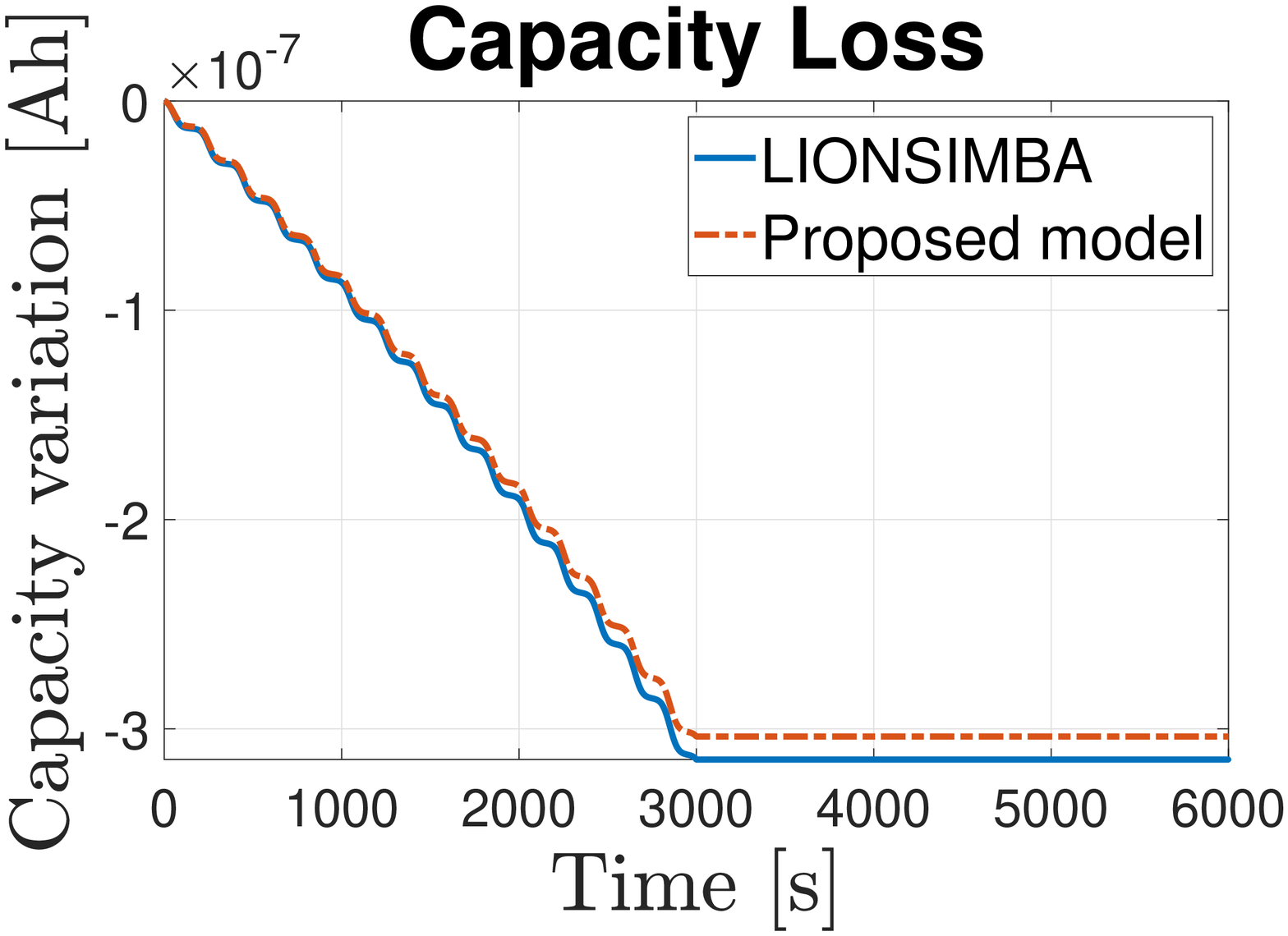}}
\subfigure[Cell Resistance Growth]{
\includegraphics[width=0.47\columnwidth]{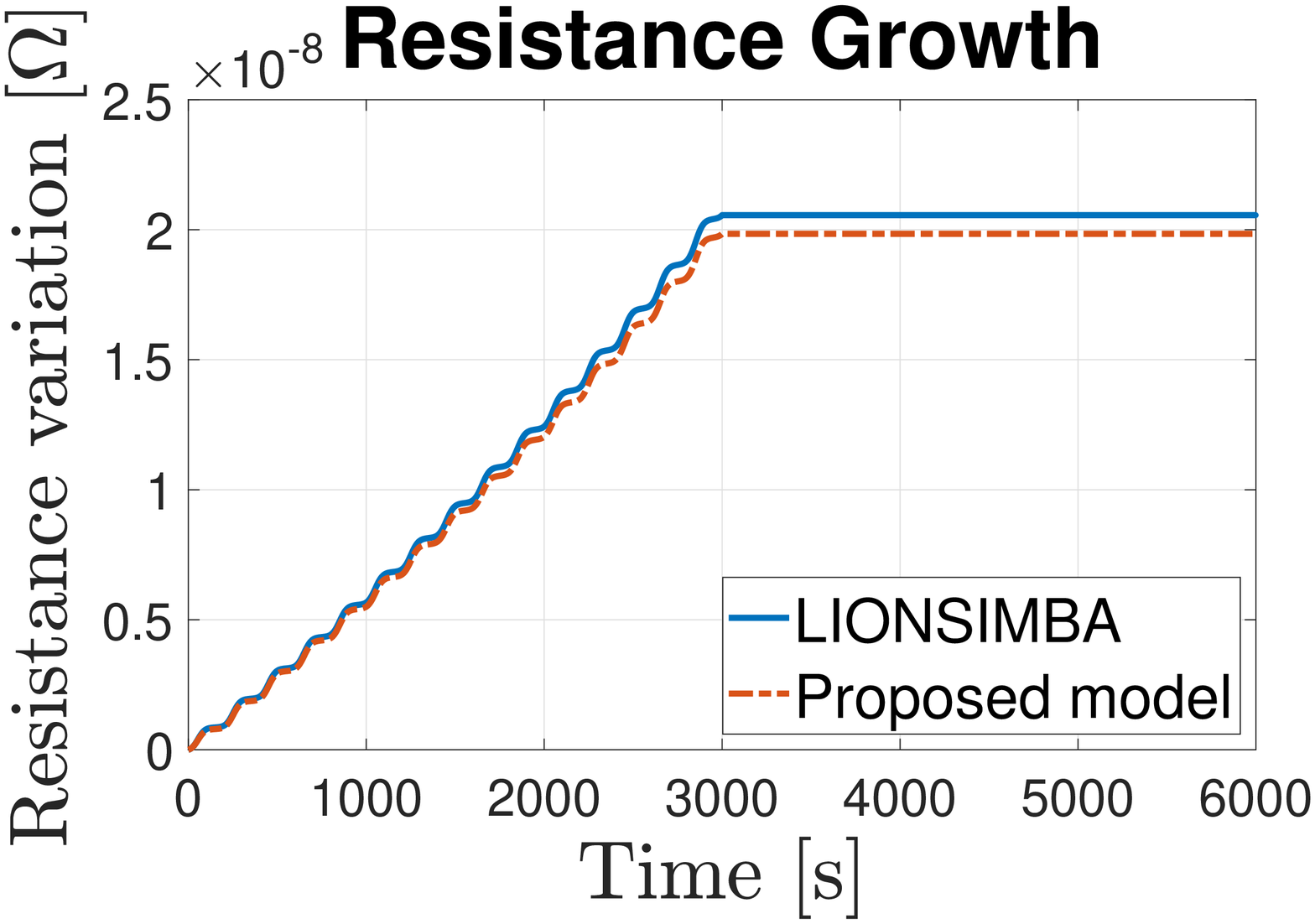}}
\caption{Ageing phenomena comparison.}
\label{fig:ageing_validation}
\end{center}
\end{figure}
Since the model described above is a modified version of the SPMe which includes also the ageing effects, in order to test its accuracy, we performed a comparison with the more detailed P2D model. In particular, for this latter, we rely on the LIONSIMBA toolbox and consider the ageing dynamics as in \cite{Santhanagopalan2006}. In order to evaluate the performance also at high-frequency, the models are excited with the sinusoidal charge-discharge input current depicted in Fig. \ref{fig:input_validation}.
 The resulting voltage, temperature and SOC are plotted respectively in Fig. \ref{fig:voltage_validation}, \ref{fig:temp_validation} and \ref{fig:soc_validation}, while the ageing effects are depicted in Fig. \ref{fig:ageing_validation}. For the sake of clarity, in this latter, only the differences from the initial values are considered. The obtained results highlight the accuracy of the proposed model.

\section{Balancing Algorithm}
\label{sec:balancing}
\textcolor{black}
In this section, we present a nonlinear MPC scheme suitable for guaranteeing a balancing-aware charging for Lithium-ion cells. In particular, in this work we consider series-connected cells. This configuration suffers from the drawback that, in the absence of a suitable supply scheme and controller, it cannot guarantee any balancing between cells exhibiting differences in terms of e.g. initial SOC, physical parameters, etc. In fact, when an input is applied to the series, the same branch current $I(t)$ flows through all the cells. 

The section is organized as follows. The optimization problem, to be solved at each time instant, is introduced in Subsection \ref{sub:optim}. The variables to be optimized are the fraction of current which flows through each one of the cells and the branch current itself. The underlying assumption is that it is possible to control  each of the former through a suitable supply scheme. In particular, Subsection \ref{sub:implem} provides an efficient, easily implementable and reliable design of the latter. Note that this is just one possible implementation which could be replaced by any other feasible one. The method explained below relies on the model described in Section \ref{sec:model}.

\subsection{Optimization Problem}
\label{sub:optim}

In the following we assume that the system is controlled by applying piece-wise constant input actions at discrete times $k$ with a sampling time $T_s$. Define as 
$\mathbf{u}(k)\in \mathbb{R}^{N+1}$
\begin{align} \label{eq:input_optim}
\mathbf{u}(k)=\left[ \delta_{1}(k) \cdots \delta_{N}(k) \ I(k) \right]^T
\end{align} 
the set of controllable inputs for the series-connected cells at each $k$, where $\delta_i(k) \in \left[ 0,1 \right]$ is the fraction of the branch current $I(k)$ flowing through the $i$-th cell during the time interval $[k, k+1)$. Note that, $I(k)$ is also considered as an optimization variable. In fact, if an adjustable current generator is available, 
the control of the branch current magnitude over time allows for better performance.

%In particular, the considered MPC formulation is based on the receding horizon idea, which enhances robustness by exploiting state feedback in order to alleviate the effects of %possible model mismatches.

\subsubsection{Cost Function}\label{sub:costfun}
In order to  consider the control objectives  as a scalar index, the following weighted cost function is considered
\begin{align}\label{eq:total_cost}
\begin{split}
J(k_0)=&\sum_{k=k_0}^{k_0+H} \sum_{i=1}^{N} \big( J_{1,i}(k) + J_{2,i}(k)+J_{3,i}(k)+J_{4}(k) \\
&+ J_{5,i}(k) + J_{6}(k)+J_{7,i}(k_0) \big )
\end{split}
\end{align}
with
\begin{subequations}
\label{eq:partial_costs}
\begin{align} 
J_{1,i}(k)&=\alpha_{1} \left(z_i(k)-\overline{z} \right)^2\label{eq:jsoc}\\
J_{2,i}(k)&= \alpha_{2} \left(\frac{T_i(k)}{T_{env}} \right)^2\label{eq:jT}\\
J_{3,i}(k)&= \alpha_{3} \delta_i(k) ^2\label{eq:jdelta}\\
J_{4}(k)&= \alpha_{4} \left(\frac{I(k)}{I_{max}}\right)^2 \label{eq:jbranch}\\
J_{5,i}(k)&= \alpha_{5} (\delta_i(k) - \delta_i(k-1))^2 \label{eq:vardelta}\\
J_{6}(k)&= \alpha_{6} \left(\frac{I(k) - I(k-1)}{I_{max}}\right)^2 \label{eq:varbranch}\\
J_{7,i}(k_0)&=\alpha_7 \left(z_i(k_0+H)-\frac{1}{N}\sum_{h=1}^N z_h(k_0+H) \right)^2\label{eq:mean}
\end{align}
\end{subequations}
where $k_0$ is the time instant at which the optimization is performed, $H$ is the prediction horizon, $T_{norm}$ is a normalization temperature which can correspond to the external environment temperature, $I_{max}$  the maximum applicable branch current and $\overline{z}$  the target SOC (usually, $\overline{z} = 1$) which the control algorithm aims to reach at the end of the charge.
The coefficients $\alpha$ need to be chosen in order to guarantee an optimal trade-off between the different objectives.
In particular, the first term \eqref{eq:jsoc} accounts for the objective of charging the cells, i.e. reaching the target state of charge $\overline{z}$.
%With the aim of mitigating the differences between the SOCs of the battery cells, which is at last the main goal of an equalization strategy, the cost \eqref{eq:javg} is introduced in order to minimize the deviations from the average state of charge at runtime, while \eqref{eq:jmeanfinal} enforces convergence at the end of the horizon. This enables to achieve an almost equalized situation even if the charge is suddenly interrupted earlier than completion.  Note that the selection of the mean battery SOC as reference is an acceptable compromise because it does not interfere with the fast charging purpose. In fact, if for instance the greatest cell SOC was chosen instead, the convergence to a common value would possibly never happen, due to the lack of penalty in the charge of the latter. On the other hand, if the lower SOC was chosen, it would lead to a slowdown of the overall charging procedure.
The term \eqref{eq:jT} aims to keep the temperature of each cell as low as possible, so as to limit safety risks as well as  degradation effects due to ageing phenomena \eqref{eq:jside}. 
Finally, the control action can be penalized through costs \eqref{eq:jdelta} and \eqref{eq:jbranch}, which account for the branch current and its fractions flowing through  each of the cells respectively. By limiting the current magnitude, energy losses and heat generation are minimized thus improving  overall battery health and efficiency. Costs \eqref{eq:vardelta} and \eqref{eq:varbranch} account for possible penalization of the input ramping rate.
The final cost \eqref{eq:mean} is introduced with the aim of speeding up the charging process. In fact, it has been noticed in simulation that without such penalty the algorithm often tends to devote most of the charging capability to a subset of the cells in the initial phase, while letting the others at a lower SOC level, so that  at the end the total charging time results  increased. This could be explained by the fact that the prediction horizon is limited and therefore it is not possible to  forecast the long-term future  behaviour of the entire system.  To this end, the terminal penalty is introduced  to consider the fact that additional charging time will be required if differences in the SOCs increase at the end of a certain prediction horizon. 

\begin{remark}
Notice that, since the proposed balancing algorithm is designed with charging capabilities only (i.e. no redistribution is allowed), the reference SOC $\overline{z}$ must be greater than all the cells initial SOC
\begin{align}
\overline{z}\geq z_i(0) \quad \forall i=1,\cdots, N
\end{align} 
\end{remark}
\subsubsection{Constraints}\label{sub:constraints}
The proper functioning of a charging system requires the consideration of some crucial physical and safety constraints. In particular, the voltage must be limited such that
\begin{subequations}\label{eq:limits}
\begin{align}\label{eq:voltage_limits}
V_i(k) \leq V_{max}
\end{align}
to avoid overvoltage which in turn could bring to safety harms and lithium plating deposition. 
Analogously, the allowed temperature should not exceed a maximum limit $T_{max}$, i.e.
\begin{align}\label{eq:temperature_limits}
T_i(k) \leq T_{max}
\end{align}
while the $SOC$ must be ensured to remain in the interval $[0,1]$ 
\begin{align}\label{eq:soc_limits}
0 \leq z_i(k) \leq 1
\end{align}
The branch current fractions $\delta_i(k)$ satisfy
\begin{align}\label{eq:delta_limits}
0 \leq \delta_i(k) \leq 1
\end{align}
by definition. Since the subject of the present work is not a redistributive charging algorithm, also a constraint on the sign of the branch current $I(k)$ must be imposed, such that (considering the charging currents as negative, according to the adopted convention)
\begin{align}\label{eq:branch_limits}
 -I_{max} \leq I(k) \leq 0 
\end{align} 
The total power supplied by the generator is bounded by the value $P_{max}(k)$ as follows
\begin{align}\label{eq:power_limits}
 -P_{max}(k) \leq \sum_{i=1}^N V_i(k) \delta_{i}(k)I(k)
\end{align}
so that a realistic scenario is considered, where the input power cannot physically be infinite. 
\end{subequations}

\subsubsection{Problem}
The resulting optimization problem, to be solved at each time $k_0$ with a prediction horizon of $H$ steps, is then the following
\begin{problem}\label{prob:problem}
Find the optimal input sequence $\mathbf{U}^*(k_0)=\left[ \mathbf{u}^*(k_0) \cdots \mathbf{u}^*(k_0+H) \right]$ such that 
\begin{align*}
\mathbf{U}^*(k_0)=\argmin_{\mathbf{U}(k_0)}J(k_0)
\end{align*}
with $J$ as in \eqref{eq:total_cost}, subject to constraints \eqref{eq:limits}.
\end{problem}
According to the receding horizon principle, only the first element $\mathbf{u}^*(k_0)$ of the obtained sequence  is applied to the system. Then, at the successive time step, the values of the relevant model states are updated according to the obtained measurements and a new optimization is performed.

\subsection{Practical Implementation}
\label{sub:implem}
The formulation of Problem \ref{prob:problem} is general in the sense that it is designed independently of the underlying power supply circuitry. In this work, with the aim of proposing a possible practical implementation,  the following supply scheme is considered.

\subsubsection{Power Supply System}
\label{sub:powsupp}
\begin{figure}[!htb]
\begin{center}
\includegraphics[scale=0.125]{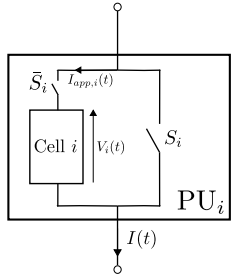}
\caption{Schematic of the $i$-th power unit (PU$_i$).}
\label{fig:pu}
\end{center}
\end{figure}
The elementary unit of the proposed supply system is a PU consisting of the $i$-th Li-Ion cell together with two power switches $S_i$ and $\bar S_i$, as shown in Fig. \ref{fig:pu}.
The two switches work in a complementary way, i.e. when $S_i$ is closed $\bar S_i$ is open and vice versa. Such design allows to fully shunt the cell when necessary. Note that, since the conduction resistance of the switches is assumed to be negligible, the terminal voltage of the $i$-th PU happens to coincide with that of the corresponding cell $V_i(t)$. The scheme described above exhibits low cost of implementation, high efficiency and the ability to be easily modularized \cite{Gallardo-Lozano2014}.  On the other hand, a possible drawback is the decay in efficiency when high currents or large  number of cells are considered. In these cases, alternative schemes can be adopted \cite{Gallardo-Lozano2014}, which still remain applicable in combination with the presented balancing-aware charging procedure. 
\begin{figure}[!htb]
\begin{center}
\includegraphics[width=0.65\columnwidth]{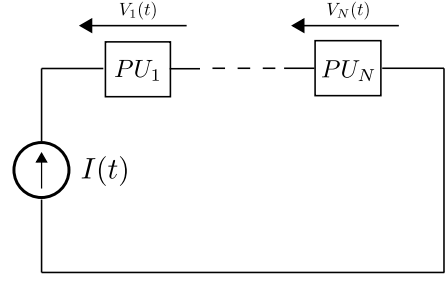}
\caption{Schematic of the power supply circuit.}
\label{fig:ps}
\end{center}
\end{figure}
\begin{figure}[!htb]
\begin{center}
\includegraphics[width=0.65\columnwidth]{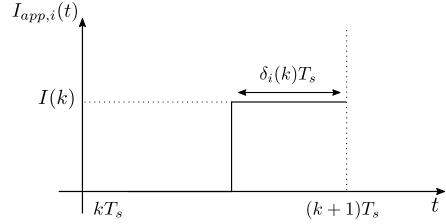}
\caption{Representation of the current on the $i$-th cell during a generic time interval.}
\label{fig:duty}
\end{center}
\end{figure}
According to Fig. \ref{fig:ps} the variable input  generator provides the branch current $I(t)$ to the whole PUs series. 
Notice that in such formulation it appears convenient to considered  the control variables $\delta_i(k)$ as duty cycles, i.e. fractions of the $k$-th time step in which the branch current $I(k)$ flows through the $i$-th cell. The interval $[kT_s,\,\, (k+1)T_s)$ is ideally divided into two parts $[kT_s,\,\,  kT_s+(1 - \delta_i(k))T_s)$ and $[kT_s + (1 - \delta_i(k))T_s,\,\, (k+1)T_s)$ as described in Fig. \ref{fig:duty}, where $I_{app,i}(t)$ is the current actually applied to the $i$-th cell over time. 
In the first part of the interval, the cell is completely bypassed (i.e. $\bar S_i$ is open and $S_i$ closed in Fig. \ref{fig:pu}), while in the second part the vice versa occurs so that the cell is charged with a current corresponding to $I(k)$.

\begin{remark}
\label{rem:oneswitch}
	Note that to avoid short circuit conditions, one must in practice make the branch current $I(t)$ vanish when all the shunting switches are closed, i.e. 
	\begin{align}
	\label{eq:oneswitch}
		I(t) = 0, \ kT_s \leq t \leq (k + 1-\max_i{\delta_i(k)})T_s, \ k \in \mathbb{N}
	\end{align}
\end{remark}

\subsubsection{Practical Considerations}
\label{sub:conservative_constraints}
\begin{figure}[!htb]
\begin{center}
\includegraphics[width=0.65\columnwidth]{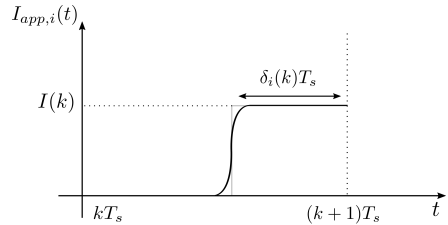}
\caption{Approximation of the current step with a sigmoid function.}
\label{fig:duty_sigmoid}
\end{center}
\end{figure}
The Optimization Problem \ref{prob:problem} is designed considering the inputs as constant during the sampling interval. However,  the profile of the  applied current (Fig. \ref{fig:duty}) is different from that of a constant current, which in this case can only correspond to the average one $I_{avg,i}(k)=\delta_i(k)I(k)$.  This brings to wrong predictions of the future states by the MPC algorithm, thus harming optimality and constraints satisfaction if the average value is considered, especially for long prediction horizons. Due to the switched structure of the supply scheme, the problem could be naturally formulated as a mixed integer program \cite{Floudas1995}. Nevertheless, the online solution of this latter is impracticable  as it becomes computationally prohibitive even if only few cells are considered. In order to preserve the possibility of an online implementation, in this work we  address the adaptation of the proposed optimization problem to the considered supply circuitry.  Since discontinuities in the input are not easily taken into account by optimization procedures, a viable solution is here proposed  in order to approximate with sufficient and arbitrary accuracy the step current profile in Fig. \ref{fig:duty}. In particular,  a continuous sigmoid function (see Fig. \ref{fig:duty_sigmoid}) is considered for any possible control step $k$, such that
\begin{equation}
\label{eq:sigmoid}
	I_{app,i}(t) = \frac{I(k)}{1+e^{-a\left(t-kT_s-(1-\delta_i)\right)}}, \quad kT_s \leq t \leq (k+1)T_s 
\end{equation}
\begin{figure}[!htb]
\begin{center}
\subfigure[Branch Current Input]{\includegraphics[width=0.45\columnwidth]{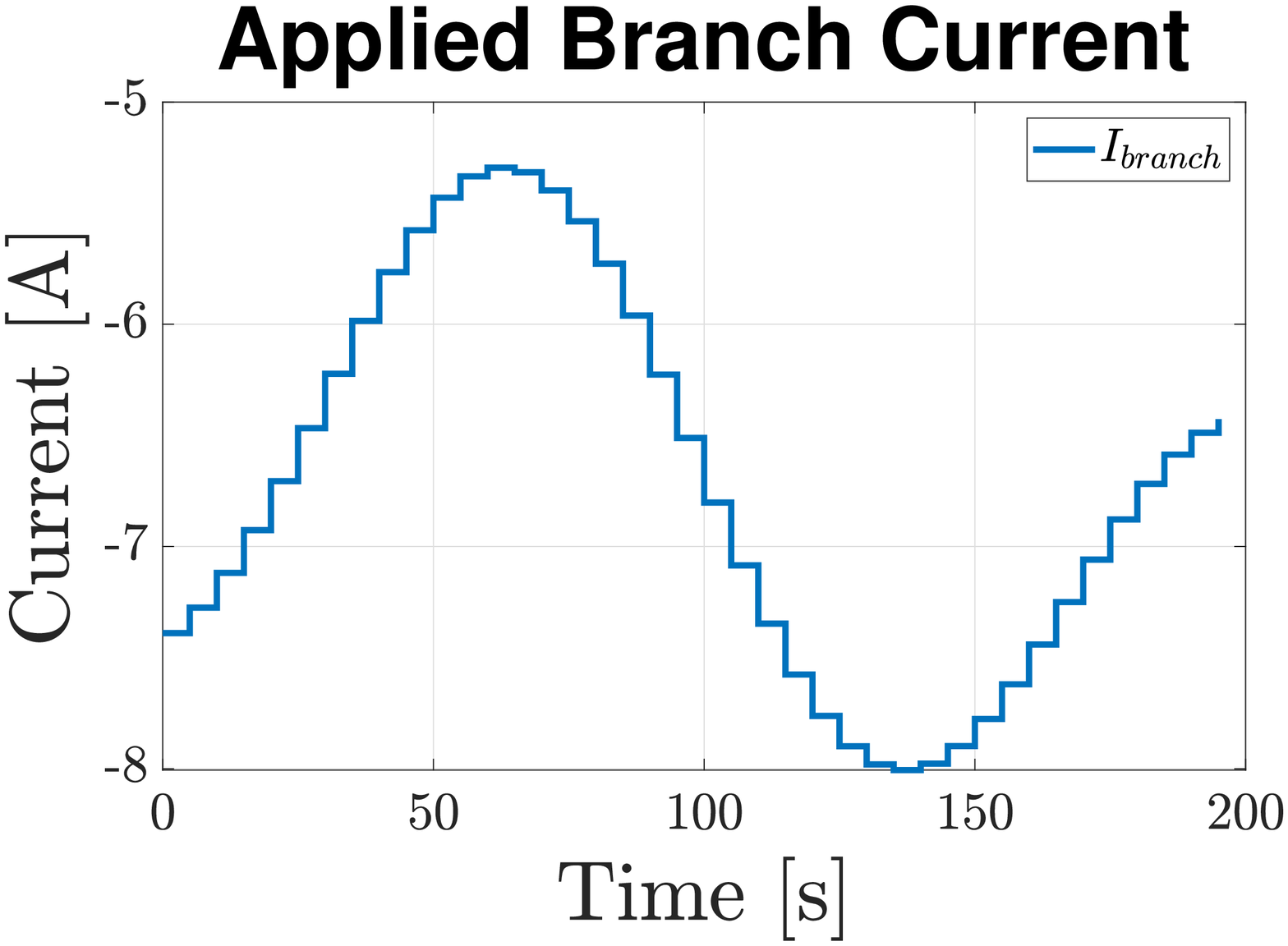}}
\subfigure[Duty Cycle Input]{\includegraphics[width=0.47\columnwidth]{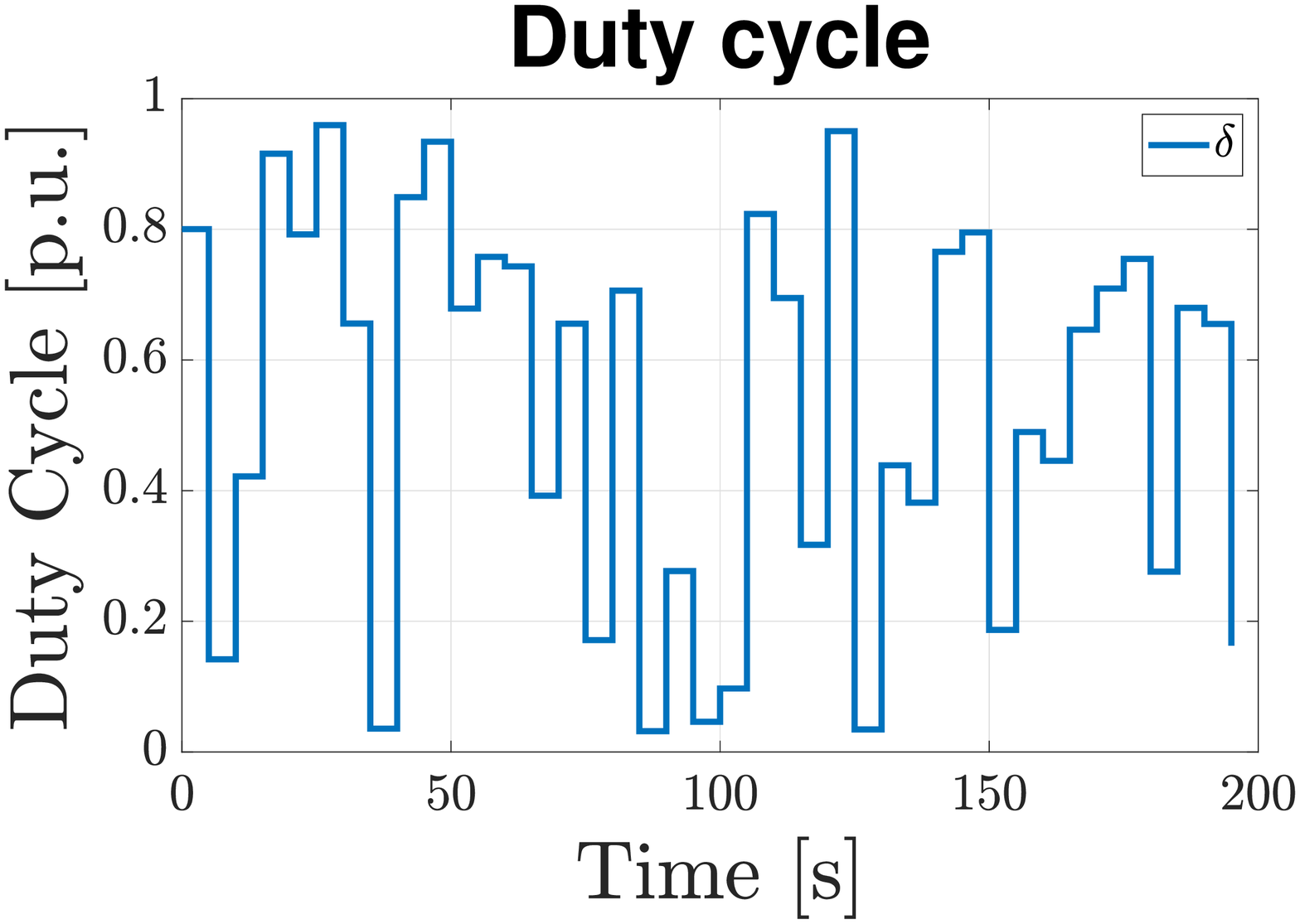}}
\caption{Time-varying inputs applied in order to evaluate the accuracy of the approximations.}
\label{fig:sigma_input}
\end{center}
\end{figure}
\begin{figure}[!htb]
\begin{center}
\includegraphics[width=0.65\columnwidth]{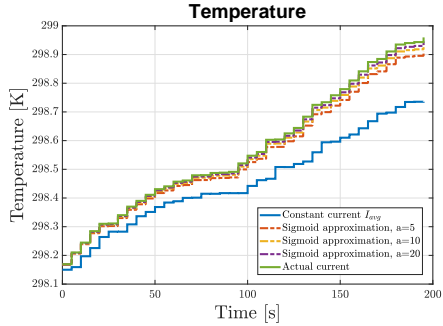}
\caption{Cell temperature approximations.}
\label{fig:temperature_sim_sig}
\end{center}
\end{figure}
\begin{figure}[!htb]
\begin{center}
\includegraphics[width=0.65\columnwidth]{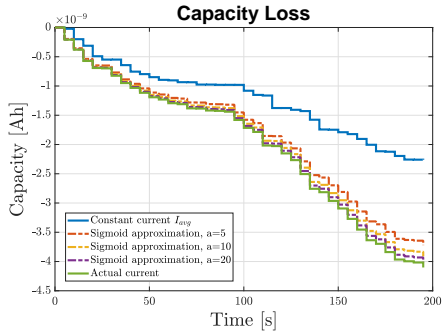}
\caption{Capacity loss  approximations.}
\label{fig:CapQ_sim_sig}
\end{center}
\end{figure}
\begin{figure}[!htb]
\begin{center}
\includegraphics[width=0.65\columnwidth]{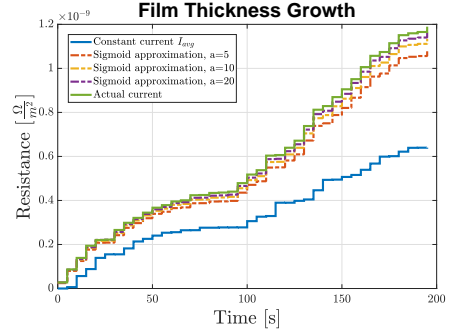}
\caption{Film growth resistance approximations.}
\label{fig:Rsei_sim_sig}
\end{center}
\end{figure}
The parameter $a$ in \eqref{eq:sigmoid} determines the slope of the sigmoid function in $kT_s + 1-\delta_i$ and therefore the accuracy of the approximation, at the expense of a possibly increased stiffness of the resulting ODEs. Simulation results, carried out for an horizon of $200s$ with the input in Fig. \ref{fig:sigma_input} are shown in Fig. \ref{fig:temperature_sim_sig}, \ref{fig:CapQ_sim_sig} and \ref{fig:Rsei_sim_sig}, where it appears evident that using the average value $I_{avg,i}(k)$ produces non negligible errors, while increasing the parameter $a$ in the sigmoid  step approximation \eqref{eq:sigmoid} enhances considerably the accuracy. Only the evolution of temperature, capacity loss and SEI resistance growth are reported, since they are the unique quantities which exhibit a nonlinear behaviour with respect to the applied current. On the other hand, no benefits in the SOC  description can be achieved through the proposed approximation,  since its dynamics is linear and sufficiently slow. Notice that the results refer to a simulation time much longer than that expected for an MPC prediction in which the real values of the considered quantities are updated every $T_s$ with the measured ones.
\begin{remark}
The possibility to control the branch current $I(k)$ through a variable current generator allows for better results. In fact, in view of the highly nonlinear relationship between temperature and input current, both the magnitude and the duration of the current applied in the second part of the control step influence the resulting behaviour. 
\end{remark}
In the following,  the possibility of a soft constraints formulation is considered in order to enhance the practical applicability of the presented equalization strategy.

\subsubsection{Constraints Softening}
\label{sub:soft}
As for the MPC formulation, a real world application could require the softening of the constraints in \eqref{eq:limits}. In particular, apart from \eqref{eq:soc_limits} and \eqref{eq:delta_limits}, which can not be relaxed outside the ranges of definition, and \eqref{eq:branch_limits}, which only regards the input, all the other constraints can be relaxed inserting slack variables $s$ such that
\begin{subequations} \label{eq:limits_relaxed}
\begin{align} \label{eq:voltage_limits_soft}
	V_i(k) \leq V_{max} + s_1(k)
\end{align}
\begin{align} \label{eq:temperature_limits_soft}
	T_i(k) \leq T_{max} + s_2(k)
\end{align}
\begin{align} \label{eq:power_limits_soft}
-P_{max}(k) - s_3(k)\leq \sum_{i=1}^N V_i(k) I_{avg,i}(k) 
\end{align}
\begin{align} \label{eq:positivity_soft}
 s_1(k) \geq 0, \quad s_2(k) \geq 0, \quad s_3(k) \geq 0
\end{align}
\end{subequations}
The total cost function $\bar J(k_0)$, at each time instant $k_0$, must be then reformulated accordingly as
\begin{align}
\label{eq:Jbar}
	\bar J(k_0) = J(k_0) + \sum_{k=k_0}^{k_0+H} \sum_{i=1}^{N} \left ( J_{s_1}(k) + J_{s_2}(k) + J_{s_3}(k) \right )
\end{align}
with $J(k)$ as in \eqref{eq:total_cost} and
\begin{subequations}
\begin{align}
	J_{s_1}(k) = \alpha_{s_1}s_1(k)
\end{align}	
\begin{align}
	J_{s_2}(k) = \alpha_{s_2}s_2(k)
\end{align}	
\begin{align}
	J_{s_3}(k) = \alpha_{s_3}s_3(k)
\end{align}
\end{subequations}

\begin{remark}
Note that softening the maximum power constraint in \eqref{eq:power_limits_soft} is feasible in practice since generators can deliver a power slightly higher than the nominal one, at least for short periods of time.
\end{remark}

\subsubsection{Practical Problem Formulation}
\label{sub:practical_opt}
The optimization problem to be solved at each step $k_0$ with a prediction horizon of $H$ steps in presence of the power supply scheme described in \ref{sub:implem} and the practical considerations therein discussed, becomes then
\begin{problem}\label{prob:practical_problem}
Find the optimal input sequence $\mathbf{U}^*(k_0)=\left[ \mathbf{u}^*(k_0) \cdots \mathbf{u}^*(k_0+H) \right]$ such that 
\begin{align*}
\mathbf{U}^*(k_0)=\argmin_{\mathbf{U}(k_0)}\bar J(k_0)
\end{align*}
with $\bar J$ as in \eqref{eq:Jbar}, subject to constraints \eqref{eq:limits}, softened as in \eqref{eq:limits_relaxed}, and \eqref{eq:oneswitch}.
\end{problem}

\section{Simulation Results}
\label{sec:simu}
In this section, the effectiveness of the proposed approach is tested in simulation. In order to obtain a realistic scenario, the proposed model is used for the control, while the more accurate LIONSIMBA simulator is considered as the real battery. In \ref{sub:testbed}, the (virtual) testbed is described in terms of adopted cells parameters and spatial configuration. At first, a standard charging protocol (namely, the CC-CV) is applied and the results are commented in \ref{sub:cccv}. Subsequently, a simple method for equalization at the end of the charge is tested in \ref{sub:ray}. At last, the same battery is charged with the proposed NMPC strategy  described in Section \ref{sec:balancing}. The improvements obtained with  such an algorithm with respect to the previous ones are highlighted in \ref{sub:results}, where a thorough discussion of the results is carried out. In particular, it is shown that the extractable capacity during the successive discharge cycle is maximized and the safety  is guaranteed by  constraints satisfaction throughout the charge. It is also evidenced that the ageing effects are  reduced. 

%% AGEING AND SOC------------------------------------------------------------------------------------------------------------------
\begin{table}[t!]
\centering
\renewcommand\arraystretch{1.3}
\begin{tabular}{c|c|c|c|c|c|c}
\toprule
\textbf{Parameter} 			& \textbf{Cell 1} 				& \textbf{Cell 2} 			& \textbf{Cell 3} 			& \textbf{Cell 4} 			& \textbf{Cell 5} 			&  \textbf{Cell 6} \\\midrule
$C^0$ [Ah]			&     $6.47$      				&  $7.62$               	& $8.83$                	& $8.68$                	& $8.46$                	& $7.22$                  \\  \midrule
      $R_{sei}^0$ [$m\Omega$] 			&     $2.09$       			& $2.36$             		& $2.22$                	& $2.36$                	& $2.12$                	& $1.76$                 \\ \midrule
      $z^0$        			&     $0.39$      				& $0.19$                	& $0.21$                	& $0.20$                	& $0.34$                	& $0.37$                 \\        
	  \bottomrule                   
\end{tabular}
\caption{Initial capacity ($C^0$), film resistance ($R_{sei}^0$) and normalized state of charge ($z^0$), values for the different cells of the pack.}
\label{tab:initial_conditions}
\end{table}
%% --------------------------------------------------------------------------------------------------------------------------------------------------

\subsection{Testbed Description}
\label{sub:testbed}
In the simulated battery pack, the considered cells are all \textit{Kokam SLPB 75106100}. All the parameters, except for the ones related to the ageing dynamics, which are taken from \cite{Santhanagopalan2006},   are   those experimentally identified in \cite{Ecker2015, Ecker2015a}.  Notice that this constitutes an appreciable feature, since in this way real commercial cells are accurately modelled. In this work the physical and thermal parameters of the cells are supposed to be equal  to the nominal ones given in \cite{Ecker2015, Ecker2015a}, with the exception of the initial capacity and SEI resistance. In fact, it is known in the literature that these latter exhibit significant variations due to both the manufacturing process and the different ageing exposure. In order to take into account such possible mismatches, the two mentioned parameters are randomly extracted as reported in Table \ref{tab:initial_conditions}, where also the initial normalized SOC is listed for each cell. 
\begin{figure}[!htb]
\begin{center}
\includegraphics[width=\figsize\columnwidth]{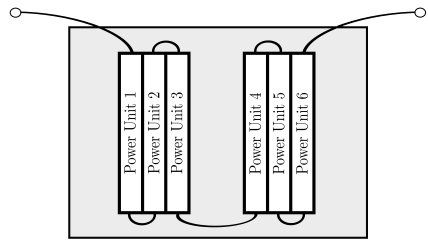}
\caption{Battery configuration. The 6 series-connected cells are grouped into two packs which are wrapped by the coolant, here indicated in light gray.}
\label{fig:battery_config}
\end{center}
\end{figure}
\begin{figure*}[!tb]
\begin{center}
\includegraphics[width=0.65\textwidth]{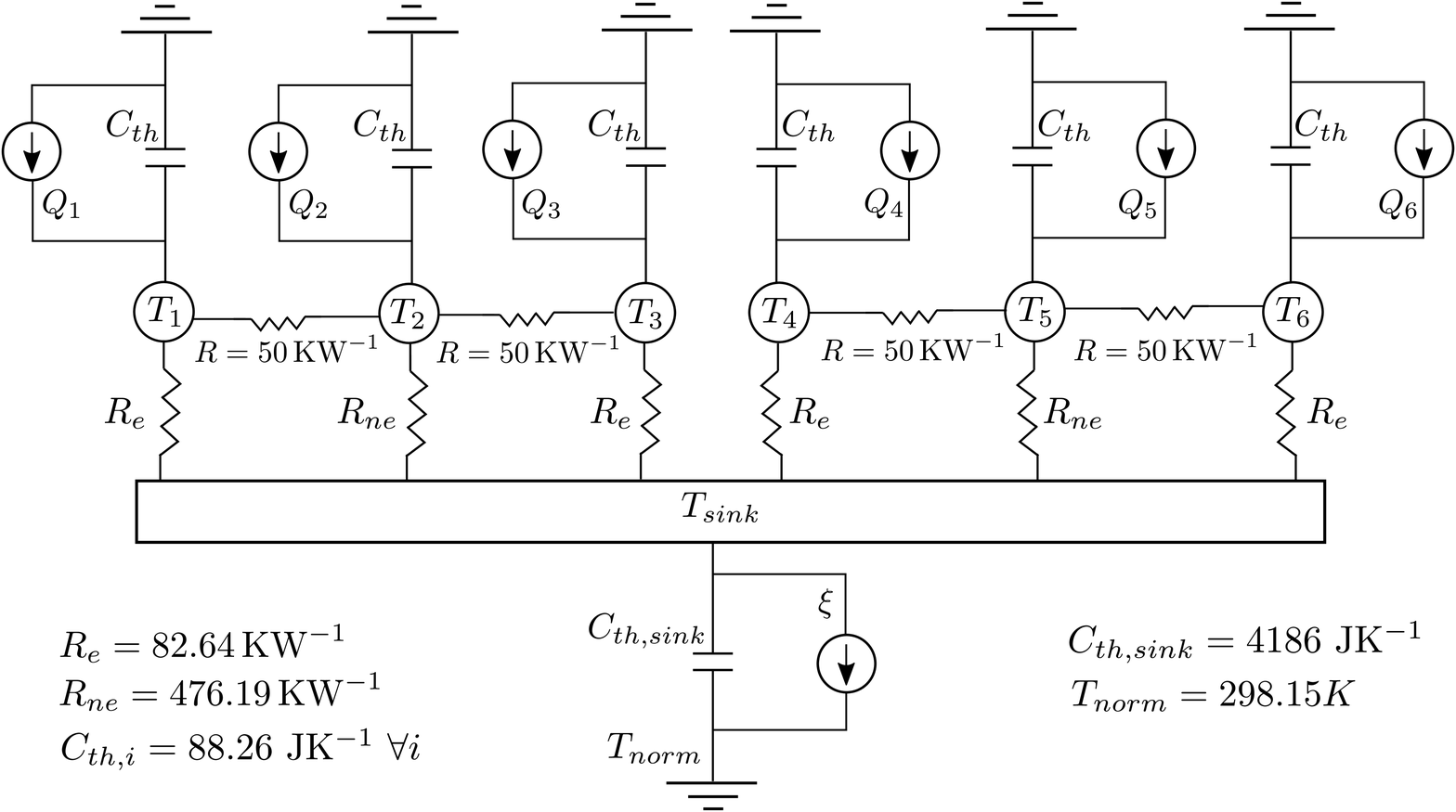}
\caption{Schematic representation of the thermal couplings using an equivalent electric circuit.  The resistances between cells are expressed in KW$^{-1}$.}
\label{fig:thermal_coupling}
\end{center}
\end{figure*}
The topology of the cells series is that depicted in Fig. \ref{fig:battery_config}, with two packs of three adjacent cells divided by a space in which the coolant is able to flow (light gray area). It also wraps the rest of the two packs, so that the most external cells have a far higher heat exchange surface compared to the inner ones.  A representation of the thermal couplings, described by means of an equivalent electric circuit, is reported in Fig. \ref{fig:thermal_coupling} along with the adopted related values.
\begin{figure}[!htb]
\begin{center}
\subfigure[Positive OCP]{\includegraphics[width=0.47\columnwidth]{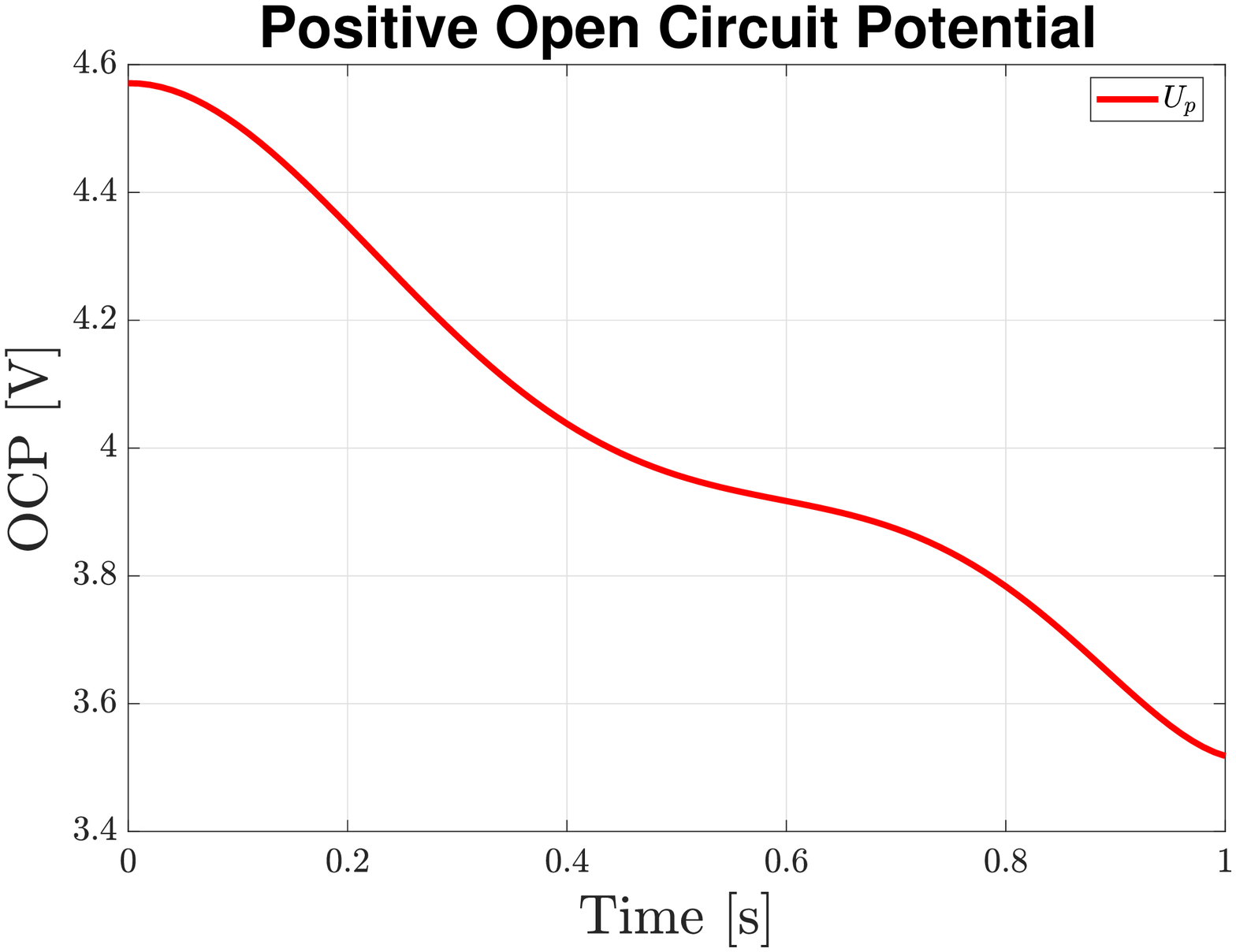}}
\subfigure[Negative OCP]{\includegraphics[width=0.47\columnwidth]{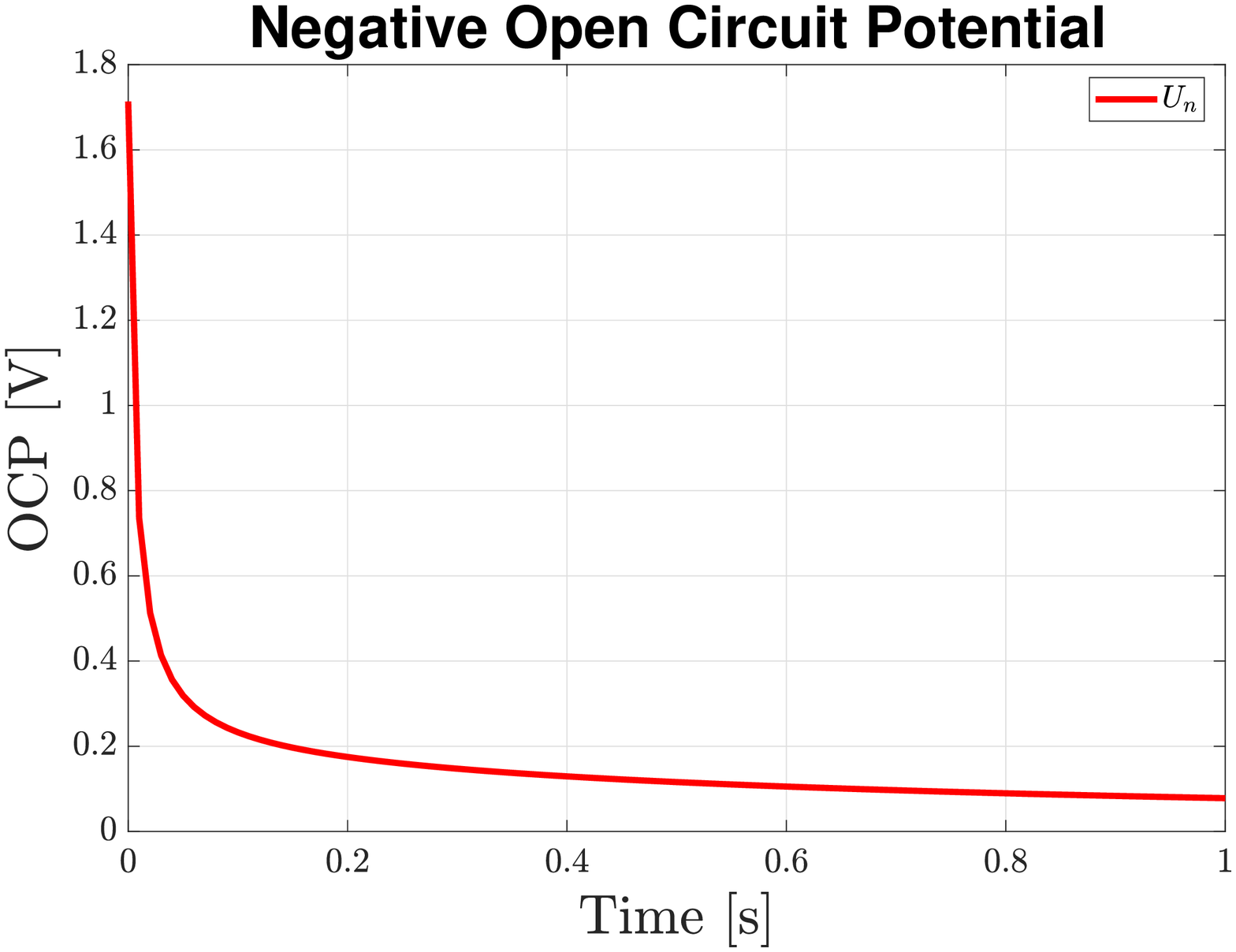}}
\caption{Nonlinear functions modelling the positive and negative OCP, respectively.}
\label{fig:ocp}
\end{center}
\end{figure}
Finally, the positive and negative open circuit potentials $\bar U_p(t)$ and $\bar U_n(t)$ for the considered cells are given by the following  functions of the surface stoichiometries $\theta_p(t)$ and $\theta_n(t)$ respectively
\begin{align}\label{eq:OCPs_fun}
\begin{split}
\bar U_p(t)=&18.45\theta_p^6(t)-40.7 \theta_p^5(t)+20.94\theta_p^4(t)\\&+8.07\theta_p^3(t)-7.837\theta_p^2(t) + 0.02414\theta_p^1(t)+4.571
\end{split}
\\
\bar U_n(t)=&    \frac{0.1261\theta_n(t)+0.00694}{\theta_n^2(t)+0.6995\theta_n(t)+0.00405}
\end{align}
Since such functions must be referred to the particular type of cell under investigation, they are obtained by a fitting procedure on the experimental data collected in \cite{Ecker2015}. The resulting relationships are provided in Fig. \ref{fig:ocp} for the readers convenience.
Finally, the temperature-dependent  electrolyte conductivity function is expressed as follows
\begin{align}\label{eq:conductivity_fun}
\begin{split}
k(\gamma_j^{[k]}(t),T(t))=&\Big(0.2667 \left(\gamma_j^{[k]}(t)\right)^3 -1.2983  \left(\gamma_j^{[k]}(t)\right)^2\\& +1.7919 \gamma_j^{[k]}(t) + 0.1726\Big)e^{\frac{-E_{a,\kappa}}{RT(t)}}
\end{split}
\end{align}
where $\gamma_j^{[k]}(t)=10^{-3}c_{e,j}^{[k]}(t)$.
Without loss of generality, in this work, the thermal power (in [$W$]) that the cooling system releases in the external environment is considered constant in the operating range, for the sake of simplicity. In particular, 
\begin{align}
	\xi(t) = 
	\begin{cases}
		5 \, &\text{if} \quad T_{sink}(t) > T_{norm} \\
		0 \, &\text{otherwise}
	\end{cases}
\end{align}

%%% AGEING AND SOC------------------------------------------------------------------------------------------------------------------
%\begin{table*}[t!]
%\centering
%\renewcommand\arraystretch{1.3}
%\begin{tabular}{c|c|c|c|c|c|c}
%\toprule
%\textbf{Parameter} 			& \textbf{Cell 1} 				& \textbf{Cell 2} 			& \textbf{Cell 3} 			& \textbf{Cell 4} 			& \textbf{Cell 5} 			&  \textbf{Cell 6} \\\midrule
%$C^0$ [Ah]			&     $6.466$      				&  $7.620$               	& $8.828$                	& $8.677$                	& $8.457$                	& $7.221$                  \\  \midrule
%      $R_{sei}^0$ [$m\Omega$] 			&     $2.09$       			& $2.36$             		& $2.22$                	& $2.36$                	& $2.12$                	& $1.76$                 \\ \midrule
%      $z^0$        			&     $0.3851$      				& $0.1868$                	& $0.2063$                	& $0.2014$                	& $0.3404$                	& $0.3684$                 \\        
%	  \bottomrule                   
%\end{tabular}
%\caption{Initial capacity ($C^0$), film resistance ($R_{sei}^0$) and normalized state of charge ($z^0$), values for the different cells of the pack.}
%\label{tab:initial_conditions}
%\end{table*}
%%% --------------------------------------------------------------------------------------------------------------------------------------------------

\subsection{Standard Charging Method}
\label{sub:cccv}
One of the most used charging protocols in industry is the well-known CC-CV \cite{Cope1999, Hussein2011}. When this protocol is applied to a set of series-connected cells, these latter are charged while keeping the total branch voltage within a specified threshold. Such limit is computed as the voltage corresponding to SOC=100\% for a nominal cell (in this case, taken as $4.15V$) multiplied by the number of cells in the series. In particular, at a first stage a constant current is applied until the limit voltage is reached and, then, such voltage is kept constant reducing the current until a specified threshold is attained.  Due to the fact that only the branch voltage is considered, some of the cells usually result in being undercharged and others overcharged. For these latter safety is at risk, while the capacity of the former is obviously not fully exploited. In order to avoid overvoltage exposure, using a voltage sensor for each of the cells, it is possible to interrupt the charging procedure as soon as one of the cells reaches its upper voltage threshold \cite{Barsukov2013}. This, however, reduces even more the overall used battery capacity. 
\begin{figure}[!htb]
\begin{center}
\subfigure[Branch current input]{\includegraphics[width=0.47\columnwidth]{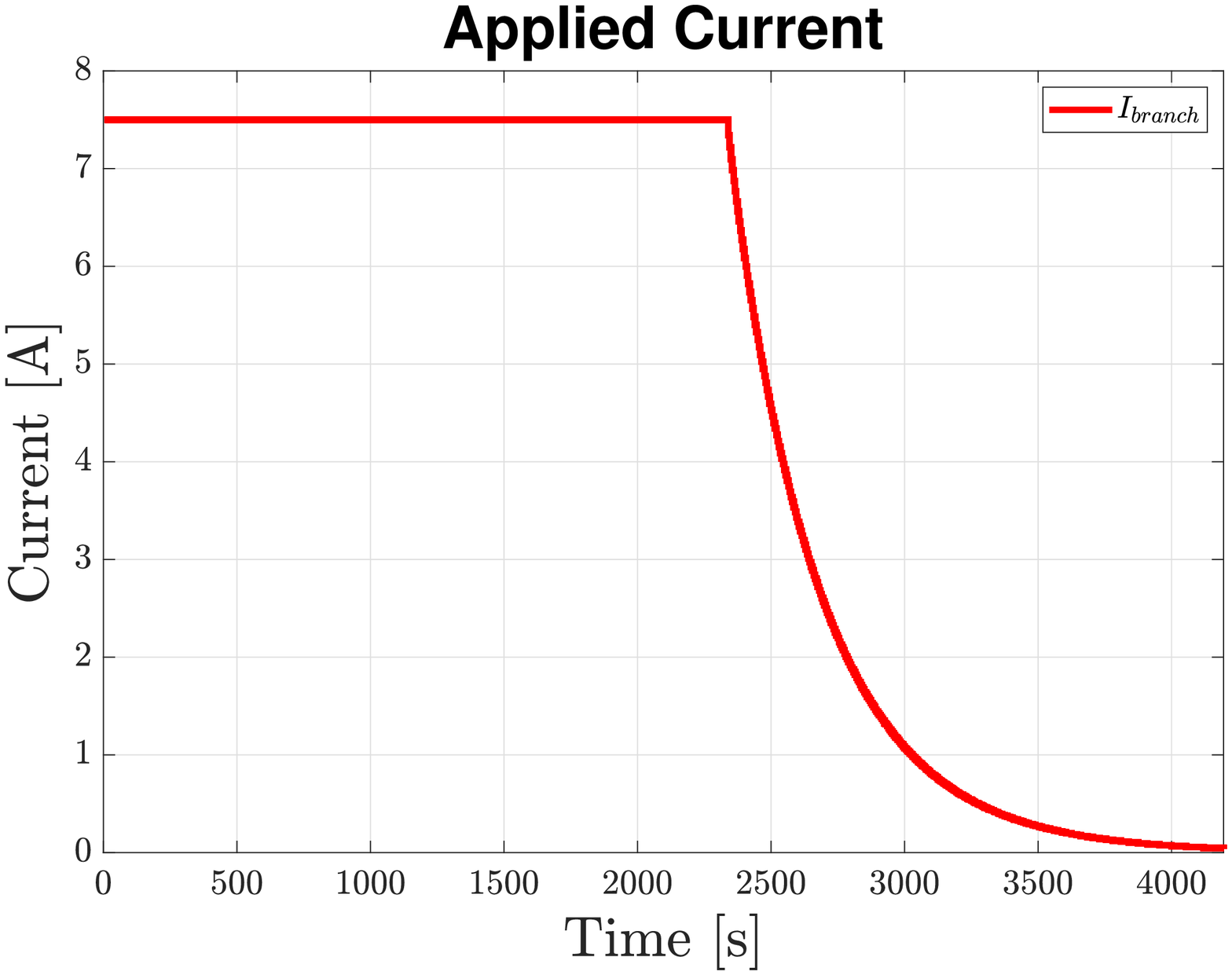}
\label{fig:cc_cv_input}}
\subfigure[Total voltage]{\includegraphics[width=0.47\columnwidth]{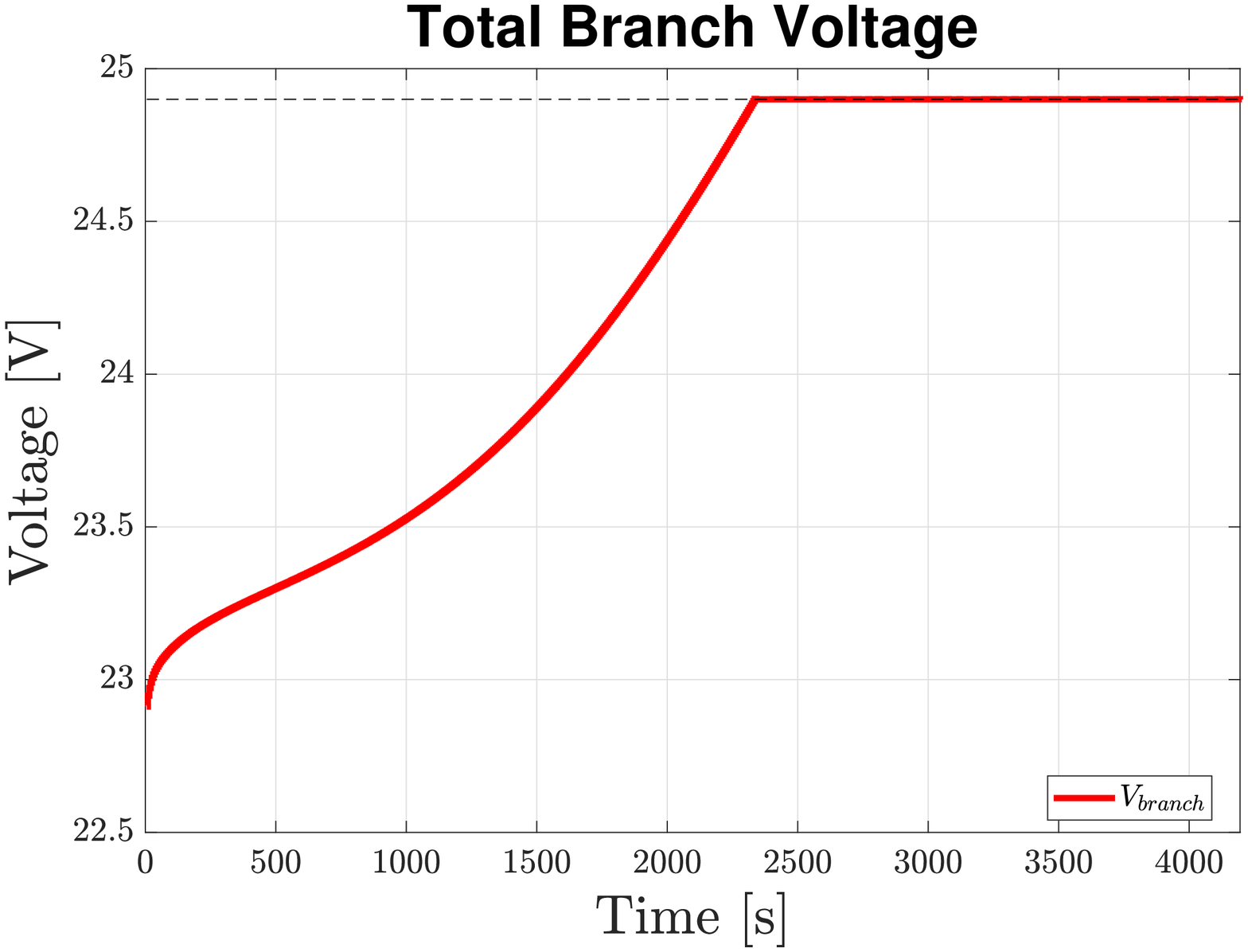}
\label{fig:cc_cv_tot_voltage}}
\caption{\textit{CC-CV charging:} Branch current  and total voltage of the cells series.}
\end{center}
\end{figure}

\begin{figure}[!htb]
\begin{center}
\includegraphics[width=\figsize\columnwidth]{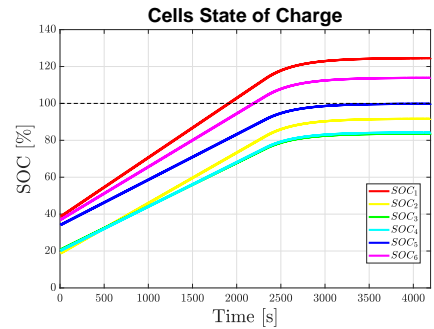}
\caption{\textit{CC-CV charging:} State of charge (SOC).}
\label{fig:cc_cv_soc}
\end{center}
\end{figure}

\begin{figure}[!htb]
\begin{center}
\includegraphics[width=\figsize\columnwidth]{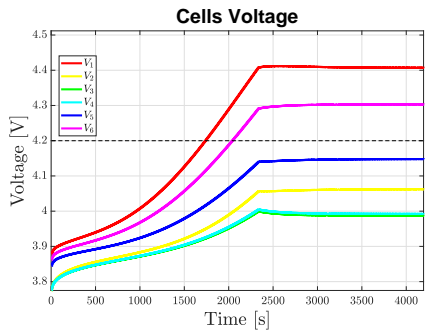}
\caption{\textit{CC-CV charging:} Voltage over the single cells.}
\label{fig:cc_cv_voltage}
\end{center}
\end{figure}
\begin{figure}[!htb]
\begin{center}
\includegraphics[width=\figsize\columnwidth]{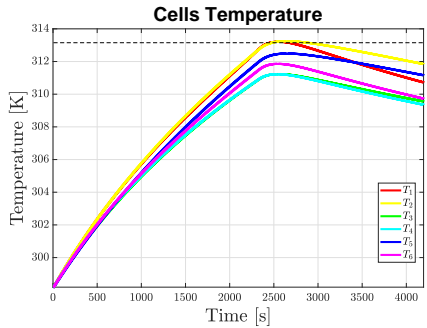}
\caption{\textit{CC-CV charging:} Temperature of the single cells.}
\label{fig:cc_cv_temperature}
\end{center}
\end{figure}

The CC-CV protocol has been applied to the testbed described in \ref{sub:testbed}. Fig. \ref{fig:cc_cv_input} reports the input branch current applied during the CC-CV. A constant current corresponding to a nominal $I_{1C}$ is firstly imposed until  the total voltage of the series reaches $24.9\, V$ (Fig. \ref{fig:cc_cv_tot_voltage}). Due to the differences in terms of total capacity, forcing the same current  through all the cells may result in the attainment of different value of SOCs at the end of the charging procedure. This could even result in the amplification of the cell differences over time,  as can be seen for the SOC in Fig. \ref{fig:cc_cv_soc}.   Moreover, the total voltage threshold is hit when some cells are well beyond their maximum limit, while others are far below full charge (Fig. \ref{fig:cc_cv_voltage}). While this can constitute a risk for safety (it can even lead to fire or explosions \cite{Wang2012}) and a sure harm to the battery health, it is also a waste of capacity, since, as made evident by Fig. \ref{fig:cc_cv_soc}, the majority of the cells are not even close to their maximum state of charge.  The temperature of the various cells, in the first part of the charge, grows mainly according to their characteristics and physical disposition (Fig. \ref{fig:cc_cv_temperature}).
In the second part of the charging, the current is decreased in order to maintain the total voltage constant. At the end of the constant-voltage procedure, SOC unbalance is still present, and therefore the total cell capacity is not exploited in a satisfactory way. During this second phase,  as expected, the temperatures decrease in response to the lower applied current. The threshold shown in Fig. \ref{fig:cc_cv_temperature} is drawn for the sake of comparison (it will be used as a constraint in the proposed NMPC algorithm). 

\begin{figure}[!htb]
\begin{center}
\includegraphics[width=\figsize\columnwidth]{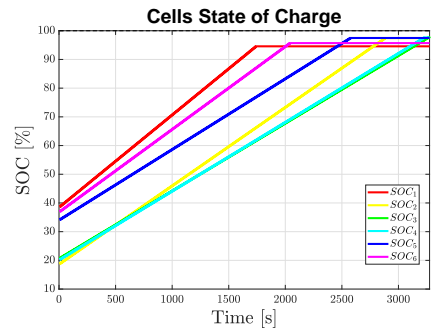}
\caption{\textit{Voltage-based charging:} State of charge (SOC).}
\label{fig:ray_soc}
\end{center}
\end{figure}
\begin{figure}[!htb]
\begin{center}
\includegraphics[width=\figsize\columnwidth]{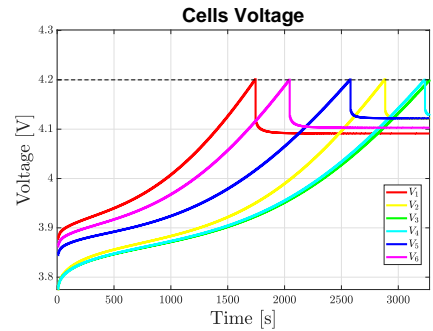}
\caption{\textit{Voltage-based charging:} Voltage over the single cells.}
\label{fig:ray_voltage}
\end{center}
\end{figure}
\begin{figure}[!htb]
\begin{center}
\includegraphics[width=\figsize\columnwidth]{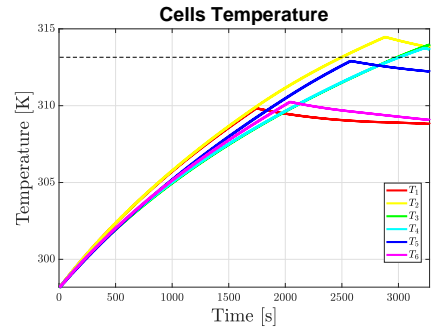}
\caption{\textit{Voltage-based charging:} Temperature of the single cells.}
\label{fig:ray_temperature}
\end{center}
\end{figure}
\subsection{Voltage-Based Method}
\label{sub:ray}
In order to obtain better performance than the CC-CV protocol, we consider also as a benchmark a simple model-less charging method which exploits the supply circuit described in Section \ref{sec:balancing}.
 Such a strategy only relies on voltage measurements and applies a constant branch current to all the cells until one of them hits its voltage threshold. Then, this latter is fully bypassed, while the charging continues for those with a voltage below the respective limit. Notice that in this case, since the constant-current phase is not followed by a constant-voltage one (a different voltage source for each cell would be required for performing the CV phase on each element of the series independently), we chose as voltage threshold the maximum allowed voltage rather than the one corresponding to $100\%$ of state of charge. 

In the carried out simulation, a constant current of $I_{1C}$ has been applied, giving rise to the results reported in Fig. \ref{fig:ray_soc}, \ref{fig:ray_voltage} and\ref{fig:ray_temperature}. As expected, the charging of each cell is interrupted as soon as the voltage threshold is hit (the maximum voltage $4.2V$, given in \cite{Ecker2015}, is considered). Therefore, the voltage limits are always respected by design. With respect to the CC-CV, it can be noticed that the SOCs are more balanced. However, they do not reach the target value of $100\%$ and  even the cell with the lowest capacity is not fully exploited. 
 \begin{figure}[!htb]
\begin{center}
\subfigure[Branch Current Input]{\includegraphics[width=0.47\columnwidth]{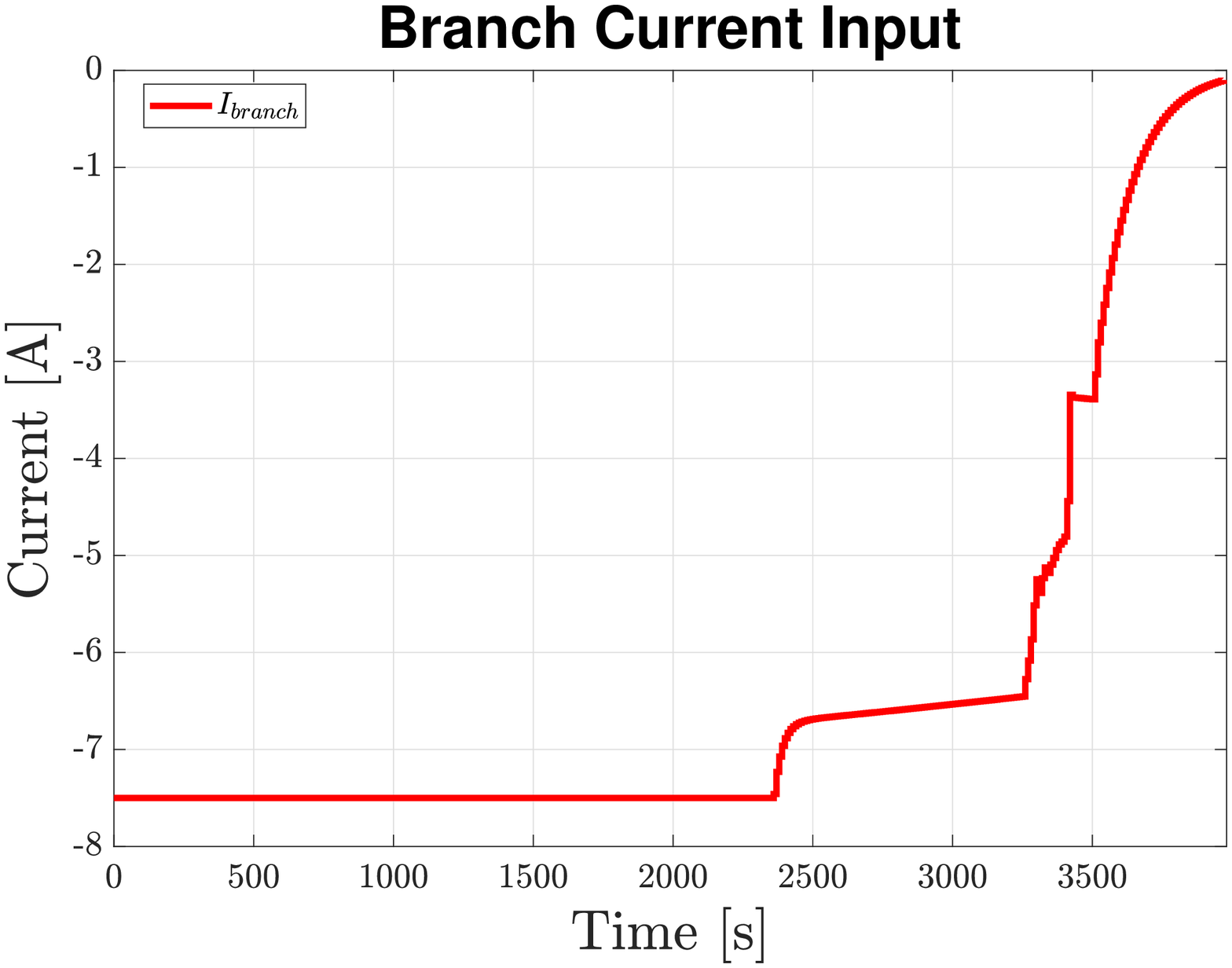} \label{fig:branchcurrent}} 
\subfigure[Duty Cycle Input]{\includegraphics[width=0.47\columnwidth]{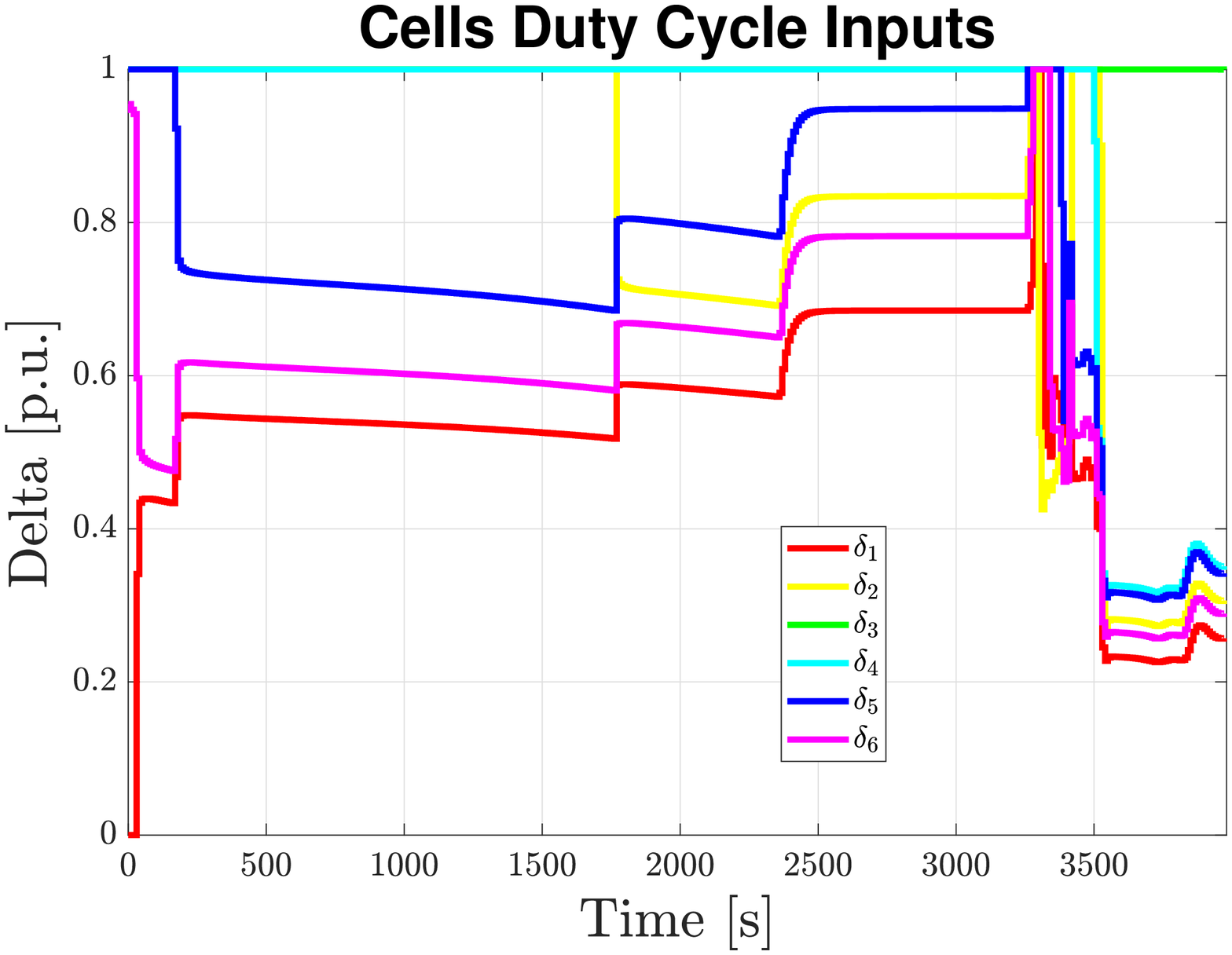}}
\caption{\textit{NMPC charging:} Optimal branch current and duty cycle inputs, where in spite of the problem highlighted in Remark \ref{rem:oneswitch}, for clarity the time intervals when the branch current is null are omitted in the plot.}\label{fig:mpc_inputs}
\end{center}
\end{figure}

\subsection{Balancing-Aware NMPC Method}
\label{sub:results}

The algorithm proposed in this Section \ref{sec:balancing} is now applied to the same battery pack considered above. 
%The results are presented from the point of view of effectiveness and then compared to that obtained with the standard CC-CV and the simple voltage-based method applied in \ref{sub:cccv} and \ref{sub:ray} respectively, in order to highlight the superiority of the optimality-based NMPC strategy. 
The values of the parameters adopted in the optimization are listed in Table \ref{tab:optimparams}. 
Notice that, due to the operating features of the considered supply scheme, the penalty on the variations of the duty cycles $\delta_i(k)$ and the branch current $I(k)$ loses its practical meaning, so that $\alpha_5 = \alpha_6 = 0$ shall be used. 
However, in the presented simulation they are taken not null with the unique purpose of enhancing the readability of the plots (for sufficiently low values of such coefficients the overall performance is not affected). The slack variables weights for the soft constraints  are taken so high that practically no relaxation takes place, i.e. $\alpha_{s_1}=\alpha_{s_2}=\alpha_{s_3}=10^{15}$.

% NMPC PARAMETERS ------------------------------------------------------------------------------------------------------------------
\begin{table}[t!]
\centering
\renewcommand\arraystretch{1.3}
\begin{tabular}{c|c|c|c|c|c}
\toprule
$\boldsymbol{N}$ 			&  $\boldsymbol{H}$ 				& $\boldsymbol{T_s}$ 			& $\boldsymbol{a}$ 			& $\boldsymbol{\overline z}$	&      $\boldsymbol{I_{max}}$    	  \\ \midrule
      $6$			&     $3$      	& $10 \ s$             & $5$           & $1$               &$7.5 \ A$	 	 \\ \toprule
     $\boldsymbol{T_{max}}$      				& $\boldsymbol{V_{max}}$                	& $\boldsymbol{P_{max}}$                	& $\boldsymbol{\alpha_1}$   &$\boldsymbol{\alpha_2}$ &
      $\boldsymbol{\alpha_3}$	         \\ \midrule
         $313.15\ K$ 				& $4.2 \ V$					& $0.75 V_{max}I_{max}$			& $10^4$ &$25$&  $0$ 							 \\ \toprule     $\boldsymbol{\alpha_4}$ 				& $\boldsymbol{\alpha_5}$					&  $\boldsymbol{\alpha_6}$			&$\boldsymbol{\alpha_7}$ & $\boldsymbol{T_{env}}$ &	$\boldsymbol{\alpha_{s_i}}, \ \forall i$	 \\ \midrule
    $1$								&  $10^{-3}$					& $10^{-3}$	&$10^5$ & $298.15$  & $10^{15}$ \\
	  \bottomrule                   
\end{tabular}
\caption{Parameters of the NMPC optimization.}
\label{tab:optimparams}
\end{table}
%% --------------------------------------------------------------------------------------------------------------------------------------------------

\begin{figure}[!htb]
\begin{center}
\includegraphics[width=\figsize\columnwidth]{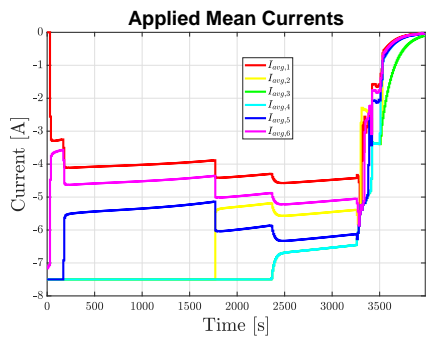}
\caption{\textit{NMPC charging:} Applied mean current to the single cells.}
\label{fig:mpc_current}
\end{center}
\end{figure}
\begin{figure}[!htb]
\begin{center}
\includegraphics[width=\figsize\columnwidth]{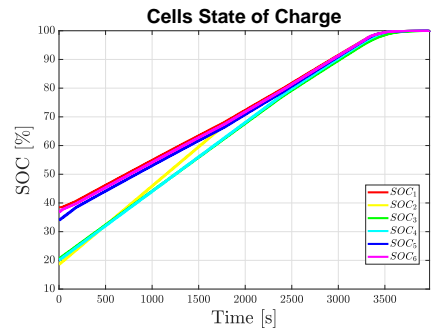}
\caption{\textit{NMPC charging:} State of charge (SOC).}
\label{fig:mpc_soc}
\end{center}
\end{figure}
\begin{figure}[!htb]
\begin{center}
\includegraphics[width=\figsize\columnwidth]{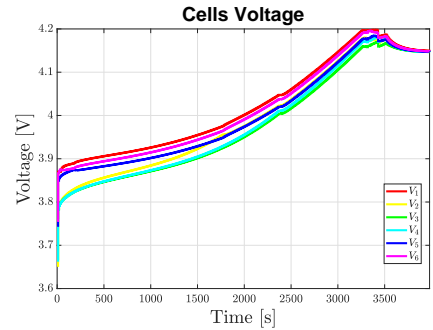}
\caption{\textit{NMPC charging:} Voltage over the single cells.}
\label{fig:mpc_voltage}
\end{center}
\end{figure}
\begin{figure}[!htb]
\begin{center}
\includegraphics[width=\figsize\columnwidth]{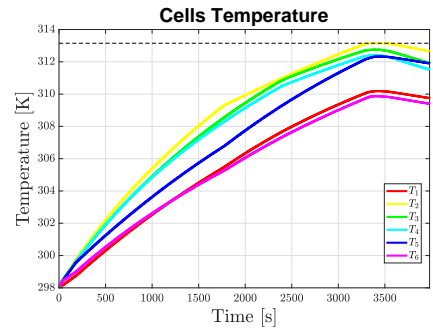}
\caption{\textit{NMPC charging:} Temperature of the single cells.}
\label{fig:mpc_temperature}
\end{center}
\end{figure}

In the following the obtained results are briefly discussed. In particular, Fig. \ref{fig:mpc_inputs} reports  branch current $I_{branch}(t)$ and duty cycles $\delta_i(t)$, which produce the average currents plotted in Fig. \ref{fig:mpc_current}.  During the whole process, the cells are charged with different mean currents, with the objective of reaching for each cell the $100\%$ of SOC while reducing the  unbalance. These goals are achieved  by a suitable tuning of $\alpha_1$ and $\alpha_7$, respectively. The inputs are obtained taking also into account the power curve of the generator. This translates into a full power supply exploitation, thus allowing for smaller and cheaper generators. As can be evidenced from Fig. \ref{fig:mpc_soc}, the SOC unbalance is reduced over time and  the procedure is stopped when all the cells are fully charged. Fig. \ref{fig:mpc_voltage} and \ref{fig:mpc_temperature} highlight the fact that both the voltage and the temperatures remain safely under the imposed limits (only the cell 2 hits and keeps the temperature threshold near the end of the charge).  This helps reducing wear and ageing, as well as improving safety. 

A key feature of the NMPC algorithm is that, with respect to the other two tested strategies, it allows for a full exploitation of the cells at the end of the charge, i.e. the SOC of all the cells is the maximum possible according to the limits. As the main benefit of this fact, one has that the charge extractable during the successive discharge cycle is maximum. In fact, assuming to discharge the whole battery pack with a constant current of $I_{1C}$ until a  cut-off voltage  threshold  ($2.7V$) is reached by one of the cells, the charge obtainable after  the application of the voltage-based method (\ref{sub:ray}) is $6.005 Ah$, while after the NMPC it is as high as $6.36Ah$ ($6\%$ more). The same arguments do not hold when taking into account the CC-CV, since it appears evident that the extractable charge is usually higher than that of the NMPC. However, the comparison is not fair since this comes at the cost of an overcharge of some of the cells (which reach a SOC as high as $124.55\%$) with sure harm to safety.

\begin{figure}[!tb]
\begin{center}
\subfigure[\textit{CC-CV charging:} Cell Capacity Loss]{
\includegraphics[width=\figsize\columnwidth]{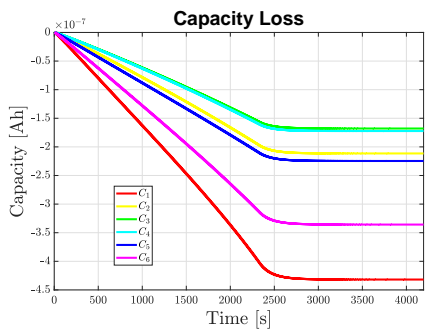} \label{fig:cc_cv_capacity}}
\subfigure[\textit{Voltage-based charging:} Cell Capacity Loss]{
\includegraphics[width=\figsize\columnwidth]{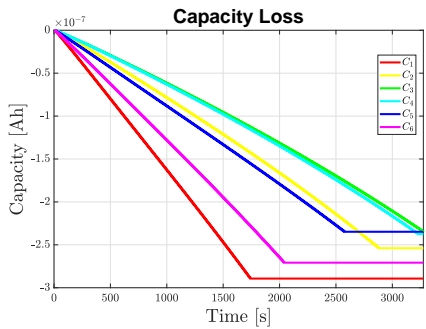}\label{fig:ray_capacity}}
\subfigure[\textit{NMPC charging:}  Cell Capacity Loss]{
\includegraphics[width=\figsize\columnwidth]{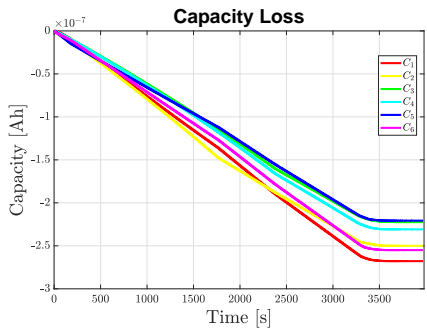}\label{fig:mpc_capacity}}
\caption{Capacity loss for the three tested algorithms.}
\label{fig:ageing_capacity}
\end{center}
\end{figure}

\begin{figure}[!tb]
\begin{center}
\subfigure[\textit{CC-CV charging:} Cell Resistance Growth]{
\includegraphics[width=\figsize\columnwidth]{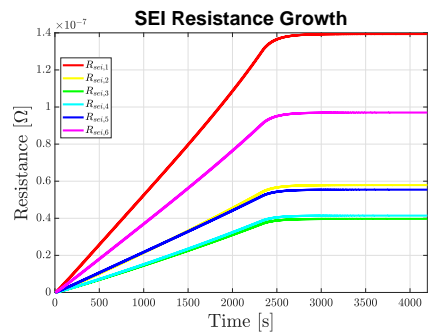}\label{fig:cc_cv_resistance}}
\subfigure[\textit{Voltage-based charging:} Cell Resistance Growth]{
\includegraphics[width=\figsize\columnwidth]{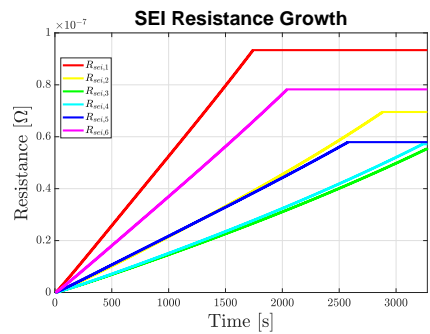}\label{fig:ray_resistance}}
\subfigure[\textit{NMPC charging:}  Cell Resistance Growth]{
\includegraphics[width=\figsize\columnwidth]{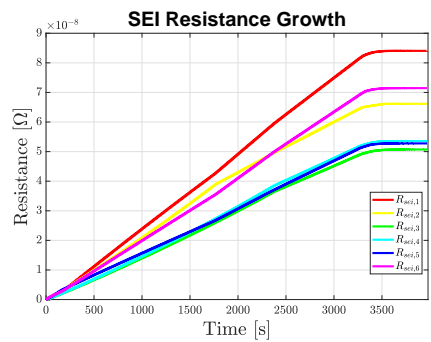}\label{fig:mpc_resistance}}
\caption{SEI resistance growth for the three tested algorithms.}
\label{fig:ageing_resistance}
\end{center}
\end{figure}
Also as for the ageing effects, the proposed algorithm outperforms the CC-CV, both from the point of view of capacity loss (\ref{fig:ageing_capacity}) and resistance growth (\ref{fig:ageing_resistance}). With respect to the voltage-based algorithm, the gain is less noticeable, but still present.

\section{Conclusions}
\label{sec:conclusions}
In the present work a NMPC  strategy has been presented for optimally charging  series-connected Lithium-ion cells, while avoiding the necessity of periodic offline balancing. A sufficiently accurate electrochemical model, suitable for control, has been formulated in Section \ref{sec:model}, which takes into account also ageing and thermal effects (comprising the coolant dynamics). This latter has then been used to establish a general NMPC framework (\ref{sub:optim}). Finally, a possible practical implementation has been considered in \ref{sub:implem}, in which the  formulation is adapted to the case of a specific implementable power supply scheme. 
The effectiveness of the proposed methodology has been validated on a virtual testbed based on the P2D electrochemical model, in which real cells parameters have been considered in order to obtain a realistic scenario. 
Simulations have shown the ability of the presented algorithm to rapidly achieve state of charge balancing while guaranteeing safety and battery health-related constraints. 
The comparison with a standard charging protocol (CC-CV) and a simple voltage-based procedure has highlighted that the  proposed  approach  outperforms  the other considered methods  from all points of view.

In order to expand the possibilities offered by nonlinear model predictive control in the context of advanced battery management systems, future research work could be devoted to solutions able to cope with parallel-of-series of cells, as well as possibly distributed architectures, in order to keep at a reasonable level the computational power requirements also in presence of thousands of cells.

\bibliography{balancing_bib}

\end{document}